\documentclass[12pt,a4paper]{article}
\pdfoutput=1
\usepackage[T1]{fontenc}
\usepackage[sc,osf]{mathpazo}
\usepackage{a4wide}  
\usepackage{amsfonts}
\usepackage{amssymb}
\usepackage{amsmath}
\usepackage{bbm}
\usepackage{booktabs} 
\usepackage{ifpdf}
\ifpdf
\usepackage[pdftex,unicode,implicit]{hyperref}
\usepackage{graphicx}
\hypersetup{%
  pdftitle    = {Supersymmetric solutions  of  SU(2)-Fayet-Iliopoulos-gauged 
N=2,d=4 supergravity}
  pdfkeywords = {supersymmetry, supergravity, gauge symmetry, solutions, vacua},
  pdfauthor   = {Tomas Ortin and Camilla Santoli},
  plainpages  = true,
  colorlinks  = true,
  citecolor   = blue,
  urlcolor    = red,
  linkcolor   = black
}
\newcommand{\hepth}[1]{{\tt
\href{http://www.arXiv.org/abs/hep-th/#1}{hep-th/#1}}}

\newcommand{\arxiv}[1]{{\tt arXiv:\href{http://www.arXiv.org/abs/#1}{#1}}}
\newcommand{\doi}[1]{{\tt DOI:\href{http://dx.doi.org/#1}{#1}}}
\else
  \usepackage[dvips]{graphicx}
  \usepackage[unicode,implicit]{hyperref}
  \newcommand{\hepth}[1]{{\tt hep-th/#1}}

  \newcommand{\arxiv}[1]{{\tt arXiv:#1}}
  \newcommand{\doi}[1]{{\tt DOI:#1}}
\fi
\makeatletter
\@addtoreset{equation}{section}
\makeatother

\begin{document}

\begin{flushright}
\small
IFT-UAM/CSIC-16-076\\
\texttt{arXiv:1609.08694 [hep-th]}\\
September 26\textsuperscript{th}, 2016\\
\normalsize
\end{flushright}

\vspace{1.5cm}

\begin{center}

{\Large {\bf Supersymmetric solutions\\[.5cm]
 of  SU$(2)$-Fayet-Iliopoulos-gauged\\[.5cm] 
$\mathcal{N}=2,d=4$ supergravity}}
 
\vspace{1.5cm}

\renewcommand{\thefootnote}{\alph{footnote}}
{\sl\large  Tom\'{a}s Ort\'{\i}n}${}^{1,}$\footnote{E-mail: {\tt Tomas.Ortin [at] csic.es}}
{\sl\large and Camilla Santoli}${}^{2,}$\footnote{E-mail: {\tt Camilla.Santoli [at] mi.infn.it}}
\setcounter{footnote}{0}
\renewcommand{\thefootnote}{\arabic{footnote}}

\vspace{1cm}

${}^{1}${\it Instituto de F\'{\i}sica Te\'orica UAM/CSIC\\
C/ Nicol\'as Cabrera, 13--15,  C.U.~Cantoblanco, E-28049 Madrid, Spain}\\ \vspace{0.3cm}

${}^{2}${\it Dipartimento di Fisica, Universit\`a di Milano, and
INFN, Sezione di Milano, Via Celoria 16, I-20133 Milano, Italy.
}

\vspace{1cm}


{\bf Abstract}

\end{center}

\begin{quotation}
  We explore the construction of supersymmetric solutions of theories of
  $\mathcal{N}=2,d=4$ supergravity with a SU$(2)$ gauging and SU$(2)$
  Fayet-Iliopoulos terms. In these theories an SU$(2)$ isometry subgroup of
  the Special-K\"ahler manifold is gauged together with a SU$(2)$ R-symmetry
  subgroup. We construct several solutions of the
  $\overline{\mathbb{CP}}{}^{3}$ quadratic model directly in four dimensions
  and of the ST$[2,6]$ model by dimensional reduction of the solutions found
  by Cariglia and Mac Conamhna in $\mathcal{N}=(1,0),d=6$ supergravity with
  the same kind of gauging. In the $\overline{\mathbb{CP}}{}^{3}$ model, we
  construct an AdS$_{2}\times$S$^{2}$ solution which is only $1/8$ BPS and an
  $\mathbb{R}\times \mathbb{H}^{3}$ solutions that also preserves $1$ of the 8
  possible supersymmetries. We show how to use dimensional reduction as in the
  ungauged case to obtain $\mathbb{R}^{n}\times $S$^{m}$ and also
  AdS$_{n}\times$S$^{m}$-type solutions (with different radii) in 5- and 4-
  dimensions from the 6-dimensional AdS$_{3}\times$S$^{3}$ solution.
\end{quotation}

\newpage
\pagestyle{plain}


\newpage

\section*{Introduction}

The study of supersymmetric solution of supergravity theories has been one of
the most fruitful areas of research in this field over the last few years
providing, for instance, backgrounds for string theory with clear spacetime
interpretation such as black holes, rings, or branes, their near-horizon
geometries, $pp$-waves etc. on which the strings can be quantized
consistently. Thus, these solutions have provided the earliest connections
between gravity solutions and 2-dimensional conformal field theories (the
superstring worldsheet theories) whose states can be counted using standard
techniques, paving the way for more general correspondences.

The supersymmetric solutions of many (classes of) supergravity theories have
been classified/characterized by now, and, therefore, the independent
variables that enter in their fields and the equations that they must obey are
well known. However, the explicit construction of these solutions can still be
a difficult problem when the equations that need to be solved are non-linear
as it is often the case in gauged supergravities, specially with non-Abelian
Yang-Mills fields. In this paper we are going to deal with this problem in the
context of $\mathcal{N}=2,d=4$ gauged supergravities. 

$\mathcal{N}=2,d=4$ supergravities admit several kinds of
gaugings:\footnote{See, for instance,
  Refs.~\cite{Andrianopoli:1996cm,Freedman:2012zz,Ortin:2015hya} for a general
  review on these theories with references to the original literature.}

\begin{enumerate}
\item One can just gauge a non-Abelian subgroup of the isometry group of the
  Special K\"ahler manifold of the complex scalars from the vector
  multiplets.\footnote{Only isometries that respect the complete Special
    Geometry structure are global symmetries of the theory and can be gauged.}
  This is the simplest possibility: it does not involve the hypermultiplets
  and trying to gauge an Abelian isometry only would have no effect since all
  the terms that would have to be added (proportional, for instance, to the
  Killing vector) vanish identically. In absence of hypermultiplets, these
  theories have been called in Refs.~\cite{Huebscher:2007hj,Hubscher:2008yz}
  $\mathcal{N}=2,d=4$ Super-Einstein-Yang-Mills (SEYM) because they are the
  simplest $\mathcal{N}=2$ supersymmetrization of the Einstein-Yang-Mills
  theories.

\item One can gauge a general subgroup of the isometry group of the
  Quaternionic K\"ahler manifold of the scalars in the
  hypermultiplets.\footnote{Only isometries that respect the Quaternionic
    K\"ahler structure are global symmetries of the theory and can be gauged.}
  Since this requires coupling to a set of gauge vector fields transforming in
  the adjoint of the gauge group and the available vectors come in
  supermultiplets that also contain scalars in a Special K\"ahler manifold,
  the gauge group must also be a subgroup of the isometry group of the Special
  K\"ahler manifold and must necessarily act on the hypermultiplets and vector
  multiplets simultaneously. It must act in the adjoint representation on the
  latter.

  This case can be considered an extension of the previous one in which the
  hypermultiplets are not mere spectators anymore. There is, however, a very
  important difference: Abelian gaugings are non-trivial in this setting in
  the Quaternionic K\"ahler sector. 

\item In absence of hypermultiplets, one can gauge the complete SU$(2)$ factor
  of the R-symmetry group (U$(2)$) or just a U$(1)$ subgroup of that SU$(2)$
  factor\footnote{The U$(1)$ factor cannot be gauged.}  by introducing what
  would be \textit{constant triholomorphic momentum maps} if there were
  hypermultiplets. These constants are usually called, respectively, SU$(2)$
  or U$(1)$ Fayet-Iliopoulos (FI) terms and the theories obtained are called
  SU$(2)$- or U$(1)$-FI-gauged $\mathcal{N}=2,d=4$ supergravities,
  respectively.

\end{enumerate}

The SU$(2)$-FI-gauged $\mathcal{N}=2,d=4$ theories can be seen as deformations
of the $\mathcal{N}=2,d=4$ SEYM theories in which the SU$(2)$ factor of the
R-symmetry group is gauged simultaneously with an SU$(2)$ subgroup of the
isometry group of the Special K\"ahler manifold. Gauging the latter is
necessary for gauging the SU$(2)$ factor of the R-symmetry group because the
global symmetry being gauged has to act on the gauge fields in the adjoint
representation and, for the gauging to respect supersymmetry, it must act on
the complete vector supermultiplets, including the scalars and this action
must, then, be an isometry of the metric.

Our goal in this paper is to search for timelike supersymmetric solutions of
this last class of gauged supergravities: SU$(2)$-FI-gauged
$\mathcal{N}=2,d=4$ supergravities with no hypermultiplets.

The timelike supersymmetric solutions of the most general $\mathcal{N}=2,d=4$
supergravities (that is: with the most general matter content and the most
general gauging) were classified/characterized in Ref.~\cite{Meessen:2012sr},
building on previous results about the supersymmetric solutions of the general
$\mathcal{N}=2,d=4$ ungauged theories with vector multiplets and
hypermultiplets \cite{Tod:1983pm,Meessen:2006tu,Huebscher:2006mr}, the
U$(1)$-FI-gauged $\mathcal{N}=2,d=4$ theories with no hypermultiplets
\cite{Caldarelli:2003pb,Cacciatori:2004rt,Cacciatori:2008ek,Klemm:2009uw} and
on the $\mathcal{N}=2,d=4$ SEYM theories,
\cite{Huebscher:2007hj,Hubscher:2008yz}.

Many solutions of the ungauged, U$(1)$-FI-gauged and SU$(2)$ SEYM theories
have been constructed in the literature but, so far, no supersymmetric
solution of SU$(2)$-FI-gauged theories is explicity known. This is due to the
complexity of the theories and of the equations that need to be solved to
construct supersymmetric solutions.  Therefore, our very first task will be to
describe carefully the structure of SU$(2)$-FI-gauged $\mathcal{N}=2,d=4$
supergravities with no hypermultiplets (Section~\ref{sec-thetheory}) and the
second will be to spell out in detail the characterization of the timelike
supersymmetric solutions of these theories found in Ref.~\cite{Meessen:2012sr}
(Section~\ref{sec-susysolutions}), showing that, according to the results of
Ref.~\cite{Louis:2016tnz}, none of them will be maximally supersymmetric.  We
will, then (Section~\ref{sec-cp3model}), consider the simplest theory that
admits an SU$(2)$ gauging, the so-called $\overline{\mathbb{CP}}^{3}$ model,
and we will perform the gauging with FI terms, constructing explicitly the
scalar potential.

In Section~\ref{sec-susycp3} we setup and try to solve by using different
methods and ansatzs the equations that the elementary building blocks of
supersymmetric solutions must satisfy in the $\overline{\mathbb{CP}}^{3}$
model. We present 3 different solutions. Finally, in
Section~\ref{sec-solutionsfromreduction} we try a different approach which is
only valid for ST$[2,n]$ models with $n\geq 6$: the authors of
Ref.~\cite{Cariglia:2004kk} constructed several timelike supersymmetric
solutions of an SU$(2)$-FI-gauged $\mathcal{N}=(1,0),d=6$ theory and, by
dimensional reduction, we can obtain solutions of the corresponding
SU$(2)$-FI-gauged $\mathcal{N}=1,d=5$ and $\mathcal{N}=2,d=4$
theories.\footnote{The solutions of SU$(2)$-FI-gauged $\mathcal{N}=1,d=5$ have
  not been received much attention, either, and, to the best of our knowledge,
  none have been presented in the literature up to this moment.}
Unfortunately, most of the solutions we obtain in this way do not have an good
asymptotically behavior (flat, AdS,..)  nor they are in general free of naked
singularities. The exception is the AdS$_{3}\times$S$^{2}$ solution which
can be obtained from the AdS$_{3}\times$S$^{3}$ one in 6 dimensions. There are
other possibilities to obtain solutions of the same type in 5 and 4 dimensions
that we explain in detail. Section~\ref{sec-conclusions} contains our
conclusions and directions for future work.

\section{SU$(2)$-FI-gauged $\mathcal{N}=2,d=4$ supergravity}
\label{sec-thetheory}

In this section we are going to review quickly the kind of theories we will be
dealing with. For more details, the reader is referred to
Refs.~\cite{Andrianopoli:1996cm,Meessen:2012sr,Ortin:2015hya}, whose
conventions we follow here. More information on the construction of these
theories can be found in Ref.~\cite{Freedman:2012zz}. 

We are considering theories of $\mathcal{N}=2,d=4$ supergravity, where the
supergravity multiplet contains the metric $g_{\mu\nu}$ and the graviphoton
vector field $A^{0}{}_{\mu}$ plus two gravitini $\psi_{I\, \mu}$, $I,J,
\ldots=1,2$, coupled to $n$ vector multiplets, each of them consisting of a
complex scalar $Z^{i}$ and a vector field $A^{i}{}_{\mu}$ plus two gaugini
$\lambda^{i\,I}$, $i=1,\cdots, n$. All the vector fields are combined into
$A^{\Lambda}{}_{\mu}$, $\Lambda,\Sigma,\ldots = 0,1,\cdots, n$. The complex
scalar parametrize a Special-K\"ahler manifold. The Special-K\"ahler
structure, which determines the K\"ahler potential $\mathcal{K}$ (and, hence,
the K\"ahler metric
$\mathcal{G}_{ij^{*}}=\partial_{i}\partial_{j^{*}}\mathcal{K}$ of the scalar
$\sigma$-model) and the \textit{period matrix}
$\mathcal{N}_{\Lambda\Sigma}(Z,Z^{*})$ that describes the coupling of the
scalars to the vector field strengths (kinetic matrices), is completely
determined by the canonical covariantly-holomorphic symplectic
section\footnote{We will also use
  \begin{equation}
  \Omega
\equiv
 e^{-\mathcal{K}/2}\mathcal{V} 
\equiv 
\left(
  \begin{array}{c}
  \mathcal{X}^{\Lambda} \\ \mathcal{X}_{\Lambda} \\
  \end{array}
\right)\, .
  \end{equation}
}
$\mathcal{V} = \left(
  \begin{smallmatrix}
  \mathcal{L}^{\Lambda} \\ \mathcal{M}_{\Lambda} \\
  \end{smallmatrix}
\right)$ or by a prepotential $\mathcal{F}$. These two objects determine
completely the ungauged theory.

The global symmetries of a theory of $\mathcal{N}=2,d=4$ supergravity coupled
to vector supermultiplets are the holomorphic isometries of the K\"ahler
metric that also preserve the rest of the Special-K\"ahler
structure\footnote{In particular, they must act as transformations of the
  symplectic group Sp$(2n+2,\mathbb{R})$ on the symplectic section and, as a
  consequence, on the period matrix.} and the R-symmetry group U$(2)$ which
only acts on the indices $I,J,K$ of the fermion fields in the fundamental
representation.  When the group of isometries that are also global symmetries
of the theory includes a non-Abelian subgroup\footnote{Abelian subgroups of
  isometries cannot be gauged in the context of $\mathcal{N}=2,d=4$ theories
  of supergravity coupled to vector supermultiplets.} which acts in the
adjoint representation on a subset of the vector supermultiplets, one can
gauge it: if the holomorphic isometries are global symmetries of the theory,
there are holomorphic Killing vectors $k_{\Lambda}(Z)$ and associated
symplectic generators of the gauge group $\mathcal{T}_{\Lambda}$ satisfying
the same Lie algebra

\begin{equation}
\label{eq:Liealgebra}
[k_{\Lambda},k_{\Sigma}]= -f_{\Lambda\Sigma}{}^{\Omega} k_{\Omega}\, ,
\hspace{1cm}
[\mathcal{T}_{\Lambda},\mathcal{T}_{\Sigma}]= 
+f_{\Lambda\Sigma}{}^{\Omega} \mathcal{T}_{\Omega}\, ,
\end{equation}

\noindent
where the $f_{\Lambda\Sigma}{}^{\Omega}$ are the structure
constants.\footnote{In this notation the generators of the gauge group carry
  the same indices as the fundamental vector fields $\Lambda$. It is
  understood that the generators, Killing vectors, structure constants
  etc.~vanish in the directions which remain ungauged. This notation is good
  enough for our purposes. A more precise (and complicated) notation would
  require the introduction of the embedding tensor to assign each generator of
  the gauge group to a gauge field.} To gauge the theory, the scalar and
vector field strengths are modified in the standard way to make them covariant
under the local transformations:\footnote{The field strengths of the fermion
  fields are also modified, but we will not be concerned with them in this
  work. See Ref.~\cite{Ortin:2015hya} for more details on this point.}

\begin{eqnarray}
\label{eq:nablazio}
\mathfrak{D}_{\mu} Z^{i} 
& = &
\partial_{\mu} Z^{i}+gA^{\Lambda}{}_{\mu} k_{\Lambda}{}^{i}\, , 
\\
& & \nonumber \\
\label{eq:Fdef}
F^{\Lambda}{}_{\mu\nu} 
& = &
2\partial_{[\mu}A^{\Lambda}{}_{\nu]}
+gf_{\Sigma\Omega}{}^{\Lambda}A^{\Sigma}{}_{[\mu}A^{\Omega}{}_{\nu]}\, .
\end{eqnarray}

\noindent
Here $g$ is the gauge coupling constant.  Furthermore, supersymmetry requires
the addition of a scalar potential which turns out to be non-negative.  The
result is the minimal $\mathcal{N}=2$ supersymmetrization of the bosonic
Einstein-Yang-Mills theory for that gauge group. These theories were called
$\mathcal{N}=2$ Super-Einstein-Yang-Mills (SEYM) theories and their timelike
supersymmetric solutions were characterized in
  Ref.~\cite{Hubscher:2008yz} and studied in
  Refs.~\cite{Huebscher:2007hj,Meessen:2008kb,Bueno:2014mea,Meessen:2015enl}.

Gauging a subgroup of the R-symmetry group seems to be a different choice,
and, indeed it is if the subgroup is Abelian (U$(1)\subset$SU$(2)$ is the only
possibility), because, as we mentioned above, Abelian holomorphic isometries
cannot be gauged in these theories. The gauging is done via Fayet-Iliopoulos
(FI) terms. The supersymmetric solutions of these theories have been
classified and studied in Refs.~\cite{Caldarelli:2003pb,Cacciatori:2008ek}.

However, when this subgroup is non-Abelian (SU$(2)$ is the only possibility,
via FI terms as well) it turns out that choice is not so different, actually:
to gauge it we need gauge vector fields transforming in the adjoint
representation of the gauge group. This implies that the whole
supermultiplets, and, in particular the complex scalars, must transform in the
adjoint representation leaving the whole Special-K\"ahler structure (and, in
particular, the K\"ahler metric) invariant. Thus, if one gauges a SU$(2)$
subgroup of the R-symmetry group one has to gauge at the same time a SU$(2)$
isometry subgroup of the global symmetry group and one can see the resulting
theory as a deformation, via FI terms, of a $\mathcal{N}=2$ SEYM theory with a
gauge group that includes a SU$(2)$ factor so that, for a subset of the vector
indices $\Lambda,\Sigma,\ldots$ that we are going to denote by the indices
$x,y,\ldots$, that only take 3 possible values, the structure constants are
those of SU$(2)$:

\begin{equation}
\label{eq:su2structureconstants}
f_{xy}{}^{z} = -\varepsilon_{xyz}\, .  
\end{equation}

These are the theories we are interested in. Their timelike supersymmetric
solutions were classified as part of the general case studied in
Ref.~\cite{Meessen:2012sr}. In the examples we will consider there will be no
other factors in the gauge group apart from the SU$(2)$ one.

Since the difference between these theories and the $\mathcal{N}=2$ SEYM
theories is the action of the gauge group on the fermions, at the bosonic
level the only difference one sees is the scalar potential, which contains
additional terms and is no longer non-negative. The scalar and vector field
strengths still take the form Eqs.~(\ref{eq:nablazio}) and
(\ref{eq:Fdef}). The bosonic action is given by

\begin{equation}
\label{eq:action}
\begin{array}{rcl}
S 
& = & 
{\displaystyle\int} d^{4}x \sqrt{|g|}
\left[R 
    +2\mathcal{G}_{ij^{*}}\mathfrak{D}_{\mu}Z^{i}\mathfrak{D}^{\mu}Z^{*\, j^{*}}
    +2\Im{\rm m}\mathcal{N}_{\Lambda\Sigma} 
    F^{\Lambda\, \mu\nu}F^{\Sigma}{}_{\mu\nu}
  \right. \\
  & & \\
  & & \left. 
    \hspace{2cm}
    -2\Re{\rm e}\mathcal{N}_{\Lambda\Sigma}  
    F^{\Lambda\, \mu\nu}\star F^{\Sigma}{}_{\mu\nu}
    -\mathbf{V}(Z,Z^{*})
  \right]\, ,
\end{array}
\end{equation}

\noindent
where the scalar potential $\mathbf{V}(Z,Z^{*})$ is given by

\begin{eqnarray}
\label{eq:potgen2}
\mathbf{V}(Z,Z^{*}) 
& = & 
-{\textstyle\frac{1}{4}}g^{2}
(\Im{\rm  
m}\mathcal{N})^{-1|\Lambda\Sigma}\mathcal{P}_{\Lambda}\mathcal{P}_{\Sigma}
\nonumber \\
& & \nonumber \\
& & 
+{\textstyle\frac{1}{2}}g^{2} 
\left(\mathcal{G}^{ij^{*}}f^{\Lambda}{}_{i}f^{*\, \Sigma}{}_{j^{*}}
-3\mathcal{L}^{*\,\Lambda} \mathcal{L}^{\Sigma}\right)
\mathsf{P}_{\Lambda}{}^{x} \mathsf{P}_{\Sigma}{}^{x}\, ,
\end{eqnarray}

\noindent
where the objects $f^{\Lambda}{}_{i}$ are the upper components of the
K\"ahler-covariant derivatives of the canonical symplectic section
$(\mathcal{D}_{i}\mathcal{V}^{M}) = \left(
  \begin{smallmatrix}
    f^{\Lambda}{}_{i} \\ h_{\Lambda\, i} \\
  \end{smallmatrix}
\right)$, $\mathcal{P}_{\Lambda}$ are the holomorphic momentum maps, and the
triholomorphic momentum maps $\mathsf{P}_{\Lambda}{}^{x}$, $x,y,\ldots=1,2,3$,
are assumed to be of the form

\begin{equation}
\label{eq:FIterms}
\mathsf{P}_{\Lambda}{}^{x}  
=
e_{\Lambda}{}^{x} \xi\, ,
\end{equation}

\noindent
for $\xi=0,1$\footnote{The role of this unphysical parameter will be to help
  us set to zero the FI terms, recovering the $\mathcal{N}=2,d=4$ SEYM
  theories.}  and constant tensors $e_{\Lambda}{}^{x}$ nonzero for $\Lambda$
in the range of the SU$(2)$ factor satisfying

\begin{equation}
\varepsilon_{xyz} e_{\Lambda}{}^{y} e_{\Sigma}{}^{z} 
= 
f_{\Lambda\Sigma}{}^{\Omega} e_{\Omega}{}^{x}\, ,  
\end{equation}

\noindent
or, taking into account Eq.~(\ref{eq:su2structureconstants}),

\begin{equation}
\varepsilon_{xy^{\prime}z^{\prime}} e_{y}{}^{y^{\prime}} e_{z}{}^{z^{\prime}} 
= 
-\varepsilon_{xyz^{\prime}} e_{z^{\prime}}{}^{x}\, .  
\end{equation}

With no loss of generality we will choose the simplest solution

\begin{equation}
\label{eq:FIterms2}
e_{x}{}^{x^{\prime}}   = - \delta_{x}{}^{x^{\prime}}\, .
\end{equation}

These constant triholomorphic momentum maps give rise to SU$(2)$
FI terms and often we will use that name for them. With this choice, the
scalar potential takes the simple form

\begin{equation}
\label{eq:potgen3}
\mathbf{V}(Z,Z^{*}) 
= 
-{\textstyle\frac{1}{4}}g^{2}
(\Im{\rm  
m}\mathcal{N})^{-1|\Lambda\Sigma}\mathcal{P}_{\Lambda}\mathcal{P}_{\Sigma}
+{\textstyle\frac{1}{2}}\xi^{2} g^{2} 
\left(\mathcal{G}^{ij^{*}}f^{x}{}_{i}f^{*\, x}{}_{j^{*}}
-3\mathcal{L}^{*\, x} \mathcal{L}^{x}\right)\, .
\end{equation}

\noindent
Observe that the first term may contain the contribution of other (necessarily
non-Abelian) gauge factors apart from the SU$(2)$ one labeled by $x,y,\ldots$
In the examples that we are going to consider we will not include that
possibility and, therefore, the sum over indices $\Lambda,\Sigma,\ldots$ will
be restricted to a sum over the SU$(2)$ indices $x,y,\ldots$

There are other differences between these theories and the SEYM ones in the
covariant derivatives of all the fermions (which now transform linearly under
the gauge group in the $I,J,\ldots$ indices)\footnote{In absence of FI terms,
  the gaugini $\lambda^{i\, I}$ transform as the scalars and vector fields in
  the same supermultiplets, on the $i,j,\ldots$ indices. The rest of the
  fermions do not transform at all.} and in the supersymmetry transformations
as well. We will not deal directly with them and, therefore, we will not
describe them here, for the sake of simplicity. All this information can be
found in the references mentioned at the beginning of this section.

\section{Timelike supersymmetric solutions}
\label{sec-susysolutions}

The timelike supersymmetric solutions of the theories introduced in the
previous section have been characterized in Ref.~\cite{Meessen:2012sr}, where
the most general gauging of these theories was considered. In this section we
are going to particularize the results obtained there to the case of the
theories we are dealing with, with only SU$(2)$ as gauge group and with the
choice of FI terms Eqs.~(\ref{eq:FIterms}) and (\ref{eq:FIterms2}).

In order to describe the form of these solutions we start by introducing an
auxiliary object $X$ with the same K\"ahler weight as the canonical symplectic
section $\mathcal{V}^{M}$ so that the quotient $\mathcal{V}^{M}/X$ has
vanishing K\"ahler weight. Then, we define two real symplectic vectors
$\mathcal{R}^{M},\mathcal{I}^{M}$

\begin{equation}
\mathcal{V}^{M}/X = \mathcal{R}^{M}+i\mathcal{I}^{M}\, .  
\end{equation}

For any model of $\mathcal{N}=2,d=4$ supergravity (or, equivalently, for any
canonical symplectic section $\mathcal{V}^{M}$) the components
$\mathcal{R}^{M}$ can, in principle, be expressed entirely in terms of the
components $\mathcal{I}^{M}$, although, in practice, this can be very hard to
do for certain models. This is often referred to as ``solving the
stabilization equations'' or as ``solving the Freudenthal duality
equations''. We will assume that this has been done and, indeed, that will be
the case in the models we will study here. Then, the symplectic product
$\mathcal{R}_{M}\mathcal{I}^{M} = \langle\, \mathcal{R} \mid \mathcal{I}\,
\rangle = \mathcal{R}_{\Lambda}\mathcal{I}^{\Lambda}
-\mathcal{R}^{\Lambda}\mathcal{I}_{\Lambda}$ is a function of the
$\mathcal{I}^{M}$ only that we call the \textit{Hesse potential}

\begin{equation}
  W(\mathcal{I}) \equiv \mathcal{R}_{M}(\mathcal{I})\mathcal{I}^{M}\, .  
\end{equation}

Now we are ready to describe the form of the fields of the timelike
supersymmetric solutions:

\begin{enumerate}
\item First of all, their metric can always be written in the
  conformastationary form\footnote{We use hats to denote differential forms.}

\begin{equation}
  ds^{2}
  = 
  e^{2U}(dt+\hat{\omega})^{2} 
  -e^{-2U}\gamma_{\underline{m}\underline{n}}dx^{m}dx^{n}\, .
\end{equation}

\noindent
The elements that enter in this expression are required to have a specific
form or satisfy certain equations:

\begin{enumerate}
\item The metric function $e^{-2U}$ is given by the Hesse potential

\begin{equation}
  e^{-2U} = W(\mathcal{I}) = \frac{1}{2|X|^{2}}\, .
\end{equation}

\item the 3-dimensional metric $\gamma_{\underline{m}\underline{n}}$ can be
  expressed in terms of Dreibein $\hat{V}^{x}$, $x=1,2,3$

\begin{equation}
  \gamma_{\underline{m}\underline{n}} 
  = 
  V^{x}{}_{\underline{m}}V^{y}{}_{\underline{n}}\delta_{xy}\, ,
\end{equation}

\noindent
and these must satisfy the equation

\begin{equation}
  d\hat{V}^{x} 
  -\xi g\epsilon^{xyz} \hat{\tilde{A}}^{y}\wedge \hat{V}^{z}
  +\hat{T}^{x}=0\, ,  
\end{equation}

\noindent
where $\hat{\tilde{A}}^{\Lambda}$ is the effective 3-dimensional gauge connection

\begin{equation}
  \label{eq:effectivegaugeconnection}
  \tilde{A}^{\Lambda}{}_{\underline{m}} 
  \equiv
  A^{\Lambda}{}_{\underline{m}}  
  +\tfrac{1}{\sqrt{2}}e^{2U}\mathcal{R}^{\Lambda}\omega_{\underline{m}}\, ,
\end{equation}

\noindent
and

\begin{eqnarray}
  \hat{T}^{x} 
  =
  \tfrac{1}{\sqrt{2}}\xi g \mathcal{I}^{y}  \hat{V}^{y}\wedge\hat{V}^{x}\, .  
\end{eqnarray}

\item The 1-form $\hat{\omega}$ satisfies the equation (in tangent 3-dimensional
  space)

  \begin{equation}
    \label{eq:do}
    (d\hat{\omega})_{xy} 
    =
    2\varepsilon_{xyz}
    \left\{
      \mathcal{I}_{M}\tilde{\mathfrak{D}}_{z}\mathcal{I}^{M}
      +\tfrac{1}{\sqrt{2}}\xi e^{-2U} \mathcal{R}^{z}
    \right\}\, ,
  \end{equation}

\noindent
where $\tilde{\mathfrak{D}}$ is the covariant derivative w.r.t.~the effective
3-dimensional gauge connection:

\begin{eqnarray}
\tilde{\mathfrak{D}}_{z}\mathcal{I}^{x} 
& = &
\partial_{z}\mathcal{I}^{x}
-g\varepsilon_{yw}{}^{x}\tilde{A}^{y}{}_{z}\mathcal{I}^{w}\, ,
\hspace{.5cm}  
\tilde{\mathfrak{D}}_{z}\mathcal{I}_{x} 
=
\partial_{z}\mathcal{I}_{x}
-g\varepsilon_{xy}{}^{w}\tilde{A}^{y}{}_{z}\mathcal{I}_{w}\,  ,
\\
& & \nonumber \\
\tilde{\mathfrak{D}}_{z}\mathcal{I}^{M} 
& = & 
\partial_{z}\mathcal{I}^{M}\, ,
\,\,\,\,\,
\mbox{when}
\,\,\,\,\,
M \neq x\, ,
\,\,\,\,\,
\mbox{(ungauged directions.)}
\end{eqnarray}

\end{enumerate}

\item The time-component of the vector fields has been gauge-fixed to

\begin{equation}
  A^{\Lambda}{}_{t} = -\tfrac{1}{\sqrt{2}}e^{2U}\mathcal{R}^{\Lambda}\, ,  
\end{equation}

\noindent
and the space components $A^{\Lambda}{}_{x}$ together with the functions
$\mathcal{I}^{M}$ are determined by the following generalization of the
Bogomol'nyi equations written again in tangent 3-dimensional space:

\begin{equation}
  \label{eq:Bogo}
  \tilde{F}^{\Lambda}{}_{xy} 
  = 
  -\tfrac{1}{\sqrt{2}}\varepsilon_{xyz} 
  \left\{
    \tilde{\mathfrak{D}}_{z}\mathcal{I}^{\Lambda}
    -\sqrt{2}\xi g\left[\mathcal{R}^{\Lambda}\mathcal{R}^{z}
      +\tfrac{1}{4} e^{-2U} 
(\Im\mathfrak{m}\mathcal{N})^{-1|\Lambda\, z}\right]
  \right\}\, ,
\end{equation}

\noindent
and

\begin{equation}
\label{eq:TimMEOMexpl}
-\tfrac{1}{\sqrt{2}}\varepsilon_{xyz}  
\tilde{\mathfrak{D}}_{x}\tilde{F}_{\Lambda\, yz} 
=  
\tfrac{1}{2}g\delta_{\Lambda}{}^{x}
\left[
g
\left(
\mathcal{I}^{x}\mathcal{I}^{y}\mathcal{I}_{y}
-\mathcal{I}_{x}\mathcal{I}^{y}\mathcal{I}^{y}
\right)
-\tfrac{1}{\sqrt{2}} \xi 
\varepsilon_{xyz}(d\hat{\omega})_{yz}
\right]\, ,
\end{equation}

\noindent
where we have \textit{defined}\footnote{There are no dual 1-forms
  $A_{\Lambda}$ in this formulation of the gauged theory.}

\begin{equation}
\label{eq:DefBeth}
\tilde{F}_{\Lambda\, xy} 
\equiv
-\tfrac{1}{\sqrt{2}}\varepsilon_{xyz}
\left\{
\tilde{\mathfrak{D}}_{z}\mathcal{I}_{\Lambda}
-\sqrt{2}g\xi
\left[ 
\mathcal{R}_{\Lambda}\mathcal{R}^{z} 
+ \tfrac{1}{4}e^{-2U}  
\Re\mathfrak{e}\mathcal{N}_{\Lambda\Gamma} 
(\Im\mathfrak{m}\mathcal{N})^{-1|\Gamma\, z}\right]
\right\}\, .
\end{equation}

\item Finally, the scalars are given by

\begin{equation}
Z^{i} 
= 
\frac{\mathcal{R}^{i}+i\mathcal{I}^{i}}{\mathcal{R}^{0}+i\mathcal{I}^{0}}\, .  
\end{equation}

\end{enumerate}

\subsection{Maximally supersymmetric vacua}

Before we start looking for explicit examples of supersymmetric solutions, it
is worth discussing the possible existence of maximally supersymmetric
solutions. According to the results of Ref.~\cite{Louis:2016tnz} the
supersymmetric solutions of these theories, if any, must be of the same kind
as those of the corresponding ungauged theories: in absence of electromagnetic
fluxes, Minkowski spacetime M$_{4}$ or anti-de Sitter spacetime
AdS$_{4}$ and, in presence of fluxes, Bertotti-Robinson spacetimes
AdS$_{2}\times$S$^{2}$ \cite{Bertotti:1959pf,Robinson:1959ev} or
Kowalski-Glikman homogeneous $pp$-wave spacetimes KG$_{4}$
\cite{KowalskiGlikman:1984wv}. Furthermore, maximally supersymmetric solutions
in gauged supergravities are characterized by the vanishing of all the fermion
shifts and of the R-symmetry connection \cite{Louis:2016tnz}.

For the $\mathcal{N}=2,d=4$ the different possibilities were analyzed in
detail in Ref.~\cite{Hristov:2009uj}. The maximally supersymmetric solutions
with zero curvature (M$_{4}$, AdS$_{2}\times$S$^{2}$ and KG$_{4}$) must have
identically vanishing triholomorphic momentum maps
$\mathsf{P}_{\Lambda}{}^{x}=0$, which is not possible in the case we are
considering. The remaining possibility is the only maximally supersymmetric
solution with negative curvature: AdS$_{4}$. The following conditions have to be
satisfied in this case:

\begin{eqnarray}
\mathsf{P}_{\Lambda}{}^{x} \mathsf{P}_{\Sigma}{}^{*\, x}
\mathcal{L}^{\Lambda} \mathcal{L}^{*\, \Sigma}
& \neq & 
0\, ,
\\
& & \nonumber \\
k_{\Lambda}{}^{i} \mathcal{L}^{*\, \Lambda}
& = & 
0\, , 
\\
& & \nonumber \\
\mathsf{P}_{\Lambda}{}^{x} f^{\Lambda}{}_{i} 
& = & 0 \, , 
\\
& & \nonumber \\
\varepsilon^{x y z} \mathsf{P}_{\Lambda}{}^{y} \mathsf{P}_{\Sigma}{}^{*\, z}
\mathcal{L}^{\Lambda} \mathcal{L}^{*\, \Sigma}
& = & 
0 \, .
\end{eqnarray}

With our choice of FI terms (\ref{eq:FIterms}),(\ref{eq:FIterms2}) these
conditions take the form 

\begin{eqnarray}
\mathcal{L}^{x} \mathcal{L}^{*\, x}
& \neq & 
0\, ,
\\
& & \nonumber \\
k_{x}{}^{i} \mathcal{L}^{*\, x}
& = & 
0\, , 
\\
& & \nonumber \\
\label{eq:msv}
f^{x}{}_{i} 
& = & 0 \, , 
\\
& & \nonumber \\
\varepsilon^{x y z} 
\mathcal{L}^{y} \mathcal{L}^{*\, z}
& = & 
0 \, .
\end{eqnarray}

Using the choice of coordinates $Z^{i}=\mathcal{X}^{i}/\mathcal{X}^{0}$ and
the gauge $\mathcal{X}^{0}=1$, it is not difficult to see, from the definition
$f^{\Lambda}{}_{i}= e^{\frac{\mathcal{K}}{2}} \mathcal{D}_i
\mathcal{X}^{\Lambda}$ that it is not possible to satisfy all the
Eqs.~(\ref{eq:msv}) at the same time.

We conclude that these theories do not admit maximally supersymmetric vacua.

\section{The SU$(2)$ gauging of the $\overline{\mathbb{CP}}^{3}$ model}
\label{sec-cp3model}

In order to search for explicit examples of supersymmetric solutions we must
specify the model of $\mathcal{N}=2,d=4$ supergravity we work with.  The
simplest example that admits an SU$(2)$ gauging is the
$\overline{\mathbb{CP}}^{3}$ model. Here we quickly review it. This model has
3 vector multiplets and the quadratic prepotential

\begin{equation}
\mathcal{F} 
=
-\tfrac{i}{4}\eta_{\Lambda\Sigma}
\mathcal{X}^{\Lambda}\mathcal{X}^{\Sigma},
\hspace{1cm}
(\eta_{\Lambda\Sigma}) = \mathrm{diag}(+---)\, .
\end{equation}

\noindent
We can define the 3 complex scalars, which parametrize a
U$(1,3)/($U$(1)\times$U$(3))$ coset space, by

\begin{equation}
Z^{i} \equiv \mathcal{X}^{i}/\mathcal{X}^{0}\, .
\end{equation}

\noindent
Adding to these $Z^{0}\equiv 1$, it is advantageous to use use $Z^{\Lambda}$
and $Z_{\Lambda}$ 

\begin{equation}
(Z^{\Lambda})
\equiv
\left(\mathcal{X}^{\Lambda}/\mathcal{X}^{0}\right) 
= 
(1,Z^{i})\, ,
\hspace{1cm}
(Z_{\Lambda})
\equiv
(\eta_{\Lambda\Sigma}Z^{\Sigma})
= 
(1,Z_{i}) 
=
(1,-Z^{i})\, .
\end{equation}

\noindent
The K\"ahler potential, the K\"ahler metric (which is the standard Bergman
metric for the symmetric space U$(1,3)/($U$(1)\times$U$(3))$
\cite{Besse:1987pua}) and its inverse in the $\mathcal{X}^{0}=1$ gauge are
given by

\begin{equation}
\label{eq:Kcpn}
\mathcal{K} 
 =  
-\log{(Z^{*\Lambda}Z_{\Lambda})}\, ,
\hspace{.4cm}
\mathcal{G}_{ij^{*}} 
= 
e^{\mathcal{K}}\left( \delta_{ij^{*}}
+e^{\mathcal{K}}Z^{*}_{i}Z_{j^{*}} \right)\, ,
\hspace{.4cm}
\mathcal{G}^{ij^{*}} 
 = 
e^{-\mathcal{K}} \left(\delta^{ij^{*}}
-Z^{i}Z^{*\, j^{*}}\right)\, ,
\end{equation}

\noindent
which implies that the complex scalars are constrained to the region 

\begin{equation}
\label{eq:CP3constraint}
0\leq \sum_{i}|Z^{i}|^{2}< 1\, .  
\end{equation}

The covariantly holomorphic symplectic section $\mathcal{V}^{M}$, its
K\"ahler-covariant derivative $\mathcal{U}_{i}=\mathcal{D}_{i}\mathcal{V}$
  and the period matrix are given by

\begin{equation}
\label{eq:Vcpn}
\mathcal{V}
=
e^{\mathcal{K}/2}
\left(
  \begin{array}{c}
  Z^{\Lambda} \\ \\ -\tfrac{i}{2} Z_{\Lambda} \\
  \end{array}
\right)\, ,
\hspace{.2cm}
\mathcal{U}_{i}
=
e^{\mathcal{K}/2}
\left(
  \begin{array}{c}
-e^{\mathcal{K}}  Z^{*}_{i}Z^{\Lambda}+\delta_{i}{}^{\Lambda} \\ 
\\ 
\tfrac{i}{2}(e^{\mathcal{K}}  Z^{*}_{i}Z_{\Lambda}-\eta_{i\Lambda} ) \\
  \end{array}
\right)\, ,
\hspace{.2cm}
\mathcal{N}_{\Lambda\Sigma}
=
\tfrac{i}{2}
\left[
\eta_{\Lambda\Sigma}
-
2\frac{Z_{\Lambda}Z_{\Sigma}}{Z^{\Gamma}Z_{\Gamma}}
\right]\, .
\end{equation}

For later use we also quote

\begin{equation}
\label{eq:periodcp3}
\Im\mathfrak{m}\, \mathcal{N}_{\Lambda\Sigma}
=
\tfrac{1}{2}
\left[
\eta_{\Lambda\Sigma}  
-\left(
\frac{Z_{\Lambda}Z_{\Sigma}}{Z^{\Gamma}Z_{\Gamma}} +\mathrm{c.c}
\right)
\right]\, ,
\hspace{.1cm}
(\Im\mathfrak{m}\, \mathcal{N})^{-1|\Lambda\Sigma}
=
2
\left[
\eta^{\Lambda\Sigma}  
-\left(
\frac{Z^{\Lambda}Z^{*\, \Sigma}}{Z^{\Gamma}Z^{*}_{\Gamma}}
+
\mathrm{c.c}
\right)
\right]\, ,
\end{equation}

\noindent
and the Hesse potential

\begin{equation}
\mathsf{W}(\mathcal{I}) 
= 
\tfrac{1}{2}\eta_{\Lambda\Sigma}\mathcal{I}^{\Lambda}\mathcal{I}^{\Sigma}
+2\eta^{\Lambda\Sigma}\mathcal{I}_{\Lambda}\mathcal{I}_{\Sigma}\, .
\end{equation}

Since the scalars parametrize the symmetric space
U$(1,3)/($U$(1)\times$U$(3))$, the metric (and, indeed, the whole model) is
invariant under global $\mathrm{U}(1,3)= \mathrm{U}(1)\times\mathrm{SU}(1,3)$
transformations. We are interested in the $\mathrm{SU}(1,3)$ subgroup whose
SO$(3)$ subgroup we are going to gauge.

The special coordinates $\mathcal{X}^{\Lambda}$ transform in the fundamental 
representation of $\mathrm{SU}(1,3)$:

\begin{equation}
\mathcal{X}^{\prime\,\Lambda}
= 
\Lambda^{\Lambda}{}_{\Sigma}\mathcal{X}^{\Sigma}\, ,
\hspace{1cm}   
\Lambda^{*\, \Gamma}{}_{\Lambda}\, \eta_{\, \Gamma\Delta}\, 
\Lambda^{\Delta}{}_{\Sigma}
=
\eta_{\Lambda\Sigma}\, ,
\end{equation}

\noindent
and, according to their definition, the complex scalars transform
non-linearly, as

\begin{equation}
\label{refsca}
Z^{ \prime\, \Lambda}
=
\frac{\Lambda^{\Lambda}{}_{\Sigma}Z^{\Sigma}}{\Lambda^{0}{}_{\Sigma}Z^{\Sigma}}\,
,
\hspace{1cm}
Z^{ \prime}{}_{\Lambda}
=
\frac{\Lambda_{\Lambda}{}^{\Sigma}Z_{\Sigma}}{\Lambda^{0}{}_{\Sigma}Z^{\Sigma}}\,
,
\,\,\,\,
\mbox{where}
\,\,\,\,
\Lambda_{\Lambda}{}^{\Sigma} \equiv
\eta_{\Lambda\Gamma}\Lambda^{\Gamma}{}_{\Omega}\eta^{\Omega \Sigma}\, .
\end{equation}

\noindent
We will use the metric $\eta_{\Lambda\Gamma}$ and its inverse to lower and
raise the indices of the SU$(1,3)$ transformations
$\Lambda^{\Lambda}{}_{\Sigma}$.

These transformations leave the K\"ahler potential invariant up to K\"ahler
transformations $\mathcal{K}^{\prime}=\mathcal{K}+f+f^{*}$ with

\begin{equation}
\label{eq:Kahlerparameter}
f(Z)
=
\log{\left(\Lambda^{0}{}_{\Sigma} Z^{\Sigma} \right)}\, ,  
\end{equation}

\noindent
which implies the exact invariance of the K\"ahler metric.

The symplectic section $\mathcal{V}^{N}$ is also left invariant by the
combined action of the symplectic transformation that gives the
embedding of the group SU$(1,3)$ in the symplectic group
Sp$(8,\mathbb{R})$

\begin{equation}
\label{eq:symplecticSU1n}
(S^{M}{}_{N})
= 
\left( 
\begin{array}{cc}
\Re\mathfrak{e}\, \Lambda^{\Lambda}{}_{\Sigma} 
& 
-2\Im\mathfrak{m}\, \Lambda^{\Lambda \Sigma} \\
& \\
\tfrac{1}{2} \Im\mathfrak{m}\, \Lambda_{\Lambda \Sigma} 
& 
\Re\mathfrak{e}\, \Lambda_{\Lambda}{}^{\Sigma} 
\end{array}
\right)\, ,
\end{equation} 

\noindent
and a K\"ahler transformation with the parameter $f(Z)$ given in
Eq.~(\ref{eq:Kahlerparameter}). This proves the invariance of the whole model
of $\mathcal{N}=2,d=4$ supergravity.

The 15 generators $T_{m}{}^{\Lambda}{}_{\Sigma}$ of $\mathfrak{su}(1,3)$,
defined by

\begin{equation}
\label{eq:infinitesimalSU1n}
\Lambda^{\Lambda}{}_{\Sigma}
\sim
\delta^{\Lambda}{}_{\Sigma}+\alpha^{m}\ T_{m}{}^{\Lambda}{}_{\Sigma},
\end{equation}

\noindent
are traceless and such that $T_{m\, \Lambda\Sigma} \equiv
\eta_{\Lambda\Gamma}\, T_{m}{}^{\Gamma}{}_{\Sigma}$ is anti-Hermitian. Then,
the corresponding $\mathfrak{sp}(1,3)$ generators, whose exponentiation gives
the matrix Eq.~(\ref{eq:symplecticSU1n}), are given by

\begin{equation}
(\mathcal{T}_{m}{}^{M}{}_{N})
=
\left( 
\begin{array}{cc}
\Re\mathfrak{e}T_{m}{}^{\Lambda}{}_{\Sigma} 
& 
-2 \Im\mathfrak{m} T_{m}{}^{\Lambda \Sigma}
\\
& \\
\tfrac{1}{2}\Im\mathfrak{m}T_{m\, \Lambda \Sigma} 
& 
\Re\mathfrak{e}T_{m\, \Lambda}{}^{\Sigma}
\\
\end{array}
\right)\, .
\end{equation}

The holomorphic Killing vectors that generate the transformations of the
scalars Eqs.~(\ref{refsca}) can be written in the form 

\begin{equation}
Z^{\prime\, \Lambda}
=
Z^{\Lambda}+\alpha^{m} k_{m}{}^{\Lambda}(Z)\, ,
\hspace{1cm}
k_{m}{}^{\Lambda}(Z)
=
T_{m}{}^{\Lambda}{}_{\Sigma}\ Z^{\Sigma}
- T_{m}{}^{0}{}_{\Omega}\  Z^{\Omega} Z^{\Lambda}\, ,   
\end{equation}

\noindent
which allows us to show easily that, if the matrices $T_{m}$ have the
commutation relations $[T_{m},T_{n}]= f_{mn}{}^{p}\, T_{p}$, where
$f_{mn}{}^{p}$ are the $\mathfrak{su}(1,3)$ structure constants, then the
commutation relations of the symplectic generators and the Lie brackets of the
holomorphic Killing vectors are given by

\begin{equation}
[\mathcal{T}_{m},\mathcal{T}_{n}]= f_{mn}{}^{p}\, \mathcal{T}_{p}\, ,
\hspace{1cm}
[k_{m},k_{n}]=-f_{mn}{}^{p}\, k_{p}\, .
\end{equation}

The holomorphic functions $\lambda_{m}(Z)$ defined through 

\begin{equation}
\mathcal{L}_{K_{m}}\mathcal{K}
=
\lambda_{m}+\lambda_{m}^{*}\, ,
\,\,\,\,\,
\mbox{where}
\,\,\,\,\,
K_{m}=k_{m}(Z)+k^{*}_{m}(Z^{*})\, , 
\end{equation}

\noindent
are given by 

\begin{equation}
\lambda_{m} 
= 
T_{m}{}^{0}{}_{\Sigma}Z^{\Sigma}\, ,  
\end{equation}

\noindent
and the holomorphic momentum maps $\mathcal{P}_{m}$, defined through
the relation 

\begin{equation}
i\mathcal{P}_{m}= k_{m}{}^{i}\partial_{i}\mathcal{K} -\lambda_{m}\, ,
\end{equation}

\noindent
are given by 

\begin{equation}
\mathcal{P}_{m} 
=
ie^{\mathcal{K}}\eta_{\Lambda\Omega}T_{m}{}^{\Lambda}{}_{\Sigma}Z^{\Sigma}Z^{*\,
  \Omega}\, .    
\end{equation}

The SU$(2)$ subgroup that we are going to gauge acts in the adjoint
representation on the special coordinates $\mathcal{X}^{i}$ and on the
physical scalars $Z^{i}$, leaving exactly invariant $\mathcal{X}^{0}$, the
prepotential and the K\"ahler potential (so $f=\lambda=0$). We are going to
use the indices $x,y,z,\cdots=1,2,3$ to denote the scalars of the gauged
directions, instead of $i,j,\cdots$. Thus, the vector fields $A^{\Lambda}$
split into $A^{0}$ and $A^{x}$, the physical scalars are $Z^{x}$, the
non-vanishing structure constants and the generators are\footnote{The indices
  $x,y,\cdots$ are raised and lowered with $\delta^{xy},\delta_{xy}$ and,
  therefore, their actual position is immaterial.}

\begin{equation}
f_{xy}{}^{z}= -\varepsilon_{xy}{}^{z}\, ,
\hspace{1cm}
T_{x}{}^{y}{}_{z} = \varepsilon_{x}{}^{y}{}_{z}\, ,
\hspace{1cm} 
(\mathcal{T}_{x}{}^{M}{}_{N})
=
\left( 
\begin{array}{cc}
\varepsilon_{x}{}^{y}{}_{z}
& 
0
\\
& \\
0
& 
\varepsilon_{xy}{}^{z}
\\
\end{array}
\right)\, ,
\end{equation}

\noindent
and the holomorphic momentum maps and Killing vectors are given by

\begin{equation}
\mathcal{P}{}_{x} = i e^{\mathcal{K}}\varepsilon_{xyz}Z^{y}Z^{*\, z}\, ,
\hspace{1cm}
k_{x}{}^{y}=\varepsilon_{x}{}^{y}{}_{z}Z^{z}\, ,
\end{equation}

\noindent
and the SU$(2)$ FI terms are given by Eqs.~(\ref{eq:FIterms}) and
(\ref{eq:FIterms2}).  Then, the gauge-covariant derivatives, vector field
strengths and scalar potential of the model
Eqs.~(\ref{eq:nablazio})-(\ref{eq:potgen3}) take the form

\begin{eqnarray}
\label{eq:nablaziocp3}
\mathfrak{D}_{\mu} Z^{x} 
& = &
\partial_{\mu} Z^{x}-g\varepsilon{}^{x}{}_{yz}A^{y}{}_{\mu}Z^{z}\, , 
\\
& & \nonumber \\
F^{0}{}_{\mu\nu} 
& = &
2\partial_{[\mu}A^{0}{}_{\nu]}
\\
& & \nonumber \\
\label{eq:Fdefcp3}
F^{x}{}_{\mu\nu} 
& = &
2\partial_{[\mu}A^{x}{}_{\nu]}
-g\varepsilon^{x}{}_{yz}A^{y}{}_{[\mu}A^{z}{}_{\nu]}\, , 
\\
& & \nonumber \\
\label{eq:potgen2cp3}
\mathbf{V}(Z,Z^{*}) 
& = & 
2g^{2} 
e^{2\mathcal{K}}
(\Re\mathfrak{e} Z^{x}\Re\mathfrak{e} Z^{x})
(\Im\mathfrak{m} Z^{y}\Im\mathfrak{m} Z^{y})
\sin^{2}\alpha
+{\textstyle\frac{1}{2}}g^{2}\xi^{2} 
\left(5 -2e^{\mathcal{K}} \right)\, ,
\end{eqnarray}

\noindent
where $\alpha$ is the angle between the 3-vectors $\Re\mathfrak{e} Z^{x}$ and 
$\Im\mathfrak{m} Z^{y}$. Observe that the first term in the potential is 
non-negative but also bounded above due to Eq.~(\ref{eq:CP3constraint}):

\begin{equation}
0 \leq 2g^{2} 
(\Re\mathfrak{e} Z^{x}\Re\mathfrak{e} Z^{x})
(\Im\mathfrak{m} Z^{y}\Im\mathfrak{m} Z^{y})
\sin^{2}\alpha \leq 2g^{2}\, ,
\end{equation}

\noindent
but the second, which is associated to the FI terms, is unbounded below
($e^{\mathcal{K}}\in (1,\infty)$):

\begin{equation}
-\infty \leq {\textstyle\frac{1}{2}}g^{2}\xi^{2} 
\left(5 -e^{\mathcal{K}} \right) \leq 2g^{2}\, .
\end{equation}

We have explored the minima of this potential and we have found that there is
a minimum when all the scalar fields vanish, when one of them vanishes, when
two of them are equal or when two of them are real, but the potential is not
negative for any of these minima and, therefore, we have not been able to find
any (necessarily non-maximally supersymmetric) AdS$_{4}$ vacuum in this
theory.

As we have already mentioned, the choice of this specific model is due to its
simplicity; in particular, its Freudenthal duality equations can easily be
solved:

\begin{equation}
\label{Freudenthal}
\mathcal{R}_{\Lambda} =\tfrac{1}{2} \eta_{\Lambda \Sigma} \mathcal{I}^{\Sigma}
\, , 
\hspace{1cm} 
\mathcal{R}^{\Lambda}=-2 \eta^{\Lambda \Sigma} \mathcal{I}_{\Sigma} \, . 
\end{equation}

\section{Timelike supersymmetric solutions of the SU$(2)$ gauged
  $\overline{\mathbb{CP}}^{3}$ model}
\label{sec-susycp3}

We just have to adapt the equations of the general recipe reviewed in
Section~\ref{sec-susysolutions} to the gauged model described in the previous
section. In particular, we use the imaginary part of period matrix
Eqs.~(\ref{eq:periodcp3}) expressed in terms of the real symplectic vectors
$\mathcal{R}^{M}$ and $\mathcal{I}^{M}$ and the solution of the Freudenthal
duality equations (\ref{Freudenthal}) to eliminate $\mathcal{R}^{M}$ from the
equations. We are also going to impose 

\begin{equation}
\mathcal{I}_{\Lambda}=0\, ,  
\end{equation}

\noindent
(so that $\mathcal{R}^{\Lambda}=0$) in order to simplify the equations. In
particular, with this choice, the form $\omega$ is closed, and we set it to
zero. The equations that remain to be solved are

\begin{eqnarray}
\label{eq:cp3eq1}
F^{0}{}_{xy} 
& = &
-\tfrac{1}{\sqrt{2}}\varepsilon_{xyz}
\left\{\partial_{z}\mathcal{I}^{0} 
+\tfrac{1}{\sqrt{2}}g\xi \mathcal{I}^{0} \mathcal{I}^{z}  \right\}\, ,
\\
& & \nonumber \\
\label{eq:cp3eq2}
F^{z}{}_{xy} 
& = &
-\tfrac{1}{\sqrt{2}}\varepsilon_{xyw}
\left\{\mathfrak{D}_{w}\mathcal{I}^{z} 
+\tfrac{1}{\sqrt{2}}g\xi\left[e^{-2U}\delta^{zw} 
+\mathcal{I}^{w} \mathcal{I}^{z}\right]  
\right\}\, ,
\\
& & \nonumber \\
\label{eq:cp3eq3}
\mathfrak{D}_{\xi} \hat{V}^{x} 
& = &
-\tfrac{1}{\sqrt{2}}g\xi\mathcal{I}^{y}\hat{V}^{y} \wedge \hat{V}^{x}\, ,  
\end{eqnarray}

\noindent
where 

\begin{equation}
\mathfrak{D}_{\xi} \hat{V}^{x} \equiv d\hat{V}^{x} 
-g\xi \varepsilon^{x}{}_{yz}\hat{A}^{y}\wedge\hat{V}^{z}\, .  
\end{equation}

\noindent
For $\xi=1$, $\mathfrak{D}_{\xi} \hat{V}^{x}=\mathfrak{D} \hat{V}^{x}$ and for
$\xi=0$, (when the FI terms vanish) $\mathfrak{D}_{\xi} \hat{V}^{x}=d
\hat{V}^{x}$ and the last equation would be solved by choosing coordinates
$\hat{V}^{x}=dx^{x}$.

The integrability condition of the last equation can be obtained by acting
with $\mathfrak{D}$ on both sides and using the Ricci identity ($\xi \neq 0$)

\begin{equation}
\mathfrak{D} \mathfrak{D}_{\xi} \hat{V}^{x} 
=
-g\xi \varepsilon^{xyz}\hat{F}^{y}\wedge\hat{V}^{z}\, .
\end{equation}

\noindent
We find, up to the overall factor $g\xi$

\begin{equation}
F^{y}{}_{xy} 
+\tfrac{1}{\sqrt{2}}\varepsilon_{xyz}
\mathfrak{D}_{z}\mathcal{I}^{y} 
=
0\, ,  
\end{equation}
 
\noindent
which is satisfied if Eq.~(\ref{eq:cp3eq2}) holds.

\subsection{Hedgehog ansatz}
\label{sec-hedgehog}

It is natural to start by looking for  spherically-symmetric solutions.  We can adopt the
\textit{hedgehog ansatz} for the gauge field $A^{x}{}_{\underline{m}}$ and the
corresponding ``Higgs field'' $\Phi^{x}$:\footnote{The signs have been chosen
  so that the equations originally obtained by Protogenov in
  Ref.~\cite{Protogenov:1977tq} coincide with those studied and used in
  Refs.~\cite{Meessen:2008kb,Bueno:2014mea,Bueno:2015wva,Meessen:2015enl}.}

\begin{equation}
\label{eq:Hedgehogansatz}
-\tfrac{1}{\sqrt{2}}\mathcal{I}^{x} = \Phi^{x}(r)= -x^{x}f(r)\, ,
\hspace{1cm}
A^{x}{}_{\underline{m}} = \varepsilon^{x}{}_{\underline{m}\underline{n}}x^{n}
h(r)\, ,  
\end{equation}

\noindent
We can also assume that the 3-dimensional metric
$\gamma_{\underline{m}\underline{n}}$ is conformally flat and choose Dreibeins
of the form

\begin{equation}
\label{eq:HAV}
V^{x}{}_{\underline{m}} = \delta^{x}{}_{\underline{m}} V(r)\, . 
\end{equation}

\noindent
We can also safely assume that 

\begin{equation}
\label{eq:HAphi0}
-\tfrac{1}{\sqrt{2}}\mathcal{I}^{0} = \Phi^{0}(r)\, .  
\end{equation}

\noindent
The ansatz for the Abelian vector field $A^{0}{}_{\underline{m}}$ cannot be
spherically symmetric: we know that the potential of the Dirac monopole is not
spherically symmetric even though the field strength is. If the unit vector
$s^{m}$ indicates the direction of the Dirac string, the Dirac monopole
potential can be written in the form 

\begin{equation}
A^{0}{}_{\underline{m}}
=
\tfrac{1}{2}p\varepsilon_{mnp}\frac{s^{n}x^{p}}{r} k(w)\, ,
\,\,\,\,
\mbox{where}
\,\,\,\,
w\equiv \frac{s^{m}x^{m}}{r}\, ,
\,\,\,\,
\mbox{and}
\,\,\,\,
k(w)= (1-w)^{-1}\, .
\end{equation}

\noindent
We can make the following ansatz  in this case:

\begin{equation}
A^{0}{}_{\underline{m}}
=
\varepsilon_{mnp}\frac{s^{n}x^{p}}{r^{2}} k(r,w)\, , \label{HAA0}
\end{equation}

\noindent
so the function $k$ can have additional dependence on $r$ (not through $w$).

Substituting this ansatz into Eqs.~(\ref{eq:cp3eq1})-(\ref{eq:cp3eq3}) we get
the following differential equations:

\begin{eqnarray}
\label{eq:proto1}
V^{-1}[2h +rh^{\prime}] -f[1+gr^{2}h]  -\tfrac{1}{2}g\xi V
\left[(\Phi^{0})^{2} -r^{2}f^{2} \right] 
& = & 
0\, ,
\\
& & \nonumber \\
\label{eq:proto2}
V^{-1}[rh^{\prime}-gr^{2}h^{2}] -gr^{2}hf +rf^{\prime} +g\xi V r^{2}f^{2}
& = & 
0\, ,
\\
& & \nonumber \\
\label{eq:proto3}
(V^{-1})^{\prime} +g\xi r[hV^{-1}-f] 
& = & 
0\, ,
\\
& & \nonumber \\
\label{eq:proto4}
x^{m}\partial_{\underline{m}}k
& = & 
0\, ,
\\
& & \nonumber \\
\label{eq:proto5}
\Phi^{0\, \prime}+ V^{-1}s^{m}\left(\frac{\partial_{\underline{m}}k}{r}
  -\frac{2x^{m}k}{r^{3}}  \right)
+g\xi r V\Phi^{0}f
& = &
0\, ,
\end{eqnarray}

\noindent
where primes indicate differentiation with respect to $r$, which is the only
argument of the functions $\Phi^{0},f,h,V$.

Eq.~(\ref{eq:proto4}) above implies that $k$ is a function of $w$ only and we
are left with

\begin{equation}
\partial_{\underline{m}}k = k^{\prime}\left(\frac{s^{m}}{r}
  -\frac{wx^{m}}{r^{2}} \right)\, ,
\end{equation}

\noindent
and

\begin{equation}
s^{m}\left(\partial_{\underline{m}}k  -\frac{2x^{m}k}{r^{2}}  \right)  
=
\frac{1}{r}\frac{d~}{dw}[(1-w^{2})k]\, .
\end{equation}

This is the only term in Eq.~(\ref{eq:proto5}) that depends on $s^{m}$ and
that dependence must disappear because the corresponding equation is
spherically symmetric. Therefore, we must require that

\begin{equation}
\frac{d~}{dw}[(1-w^{2})k] = C\, ,  
\end{equation}

\noindent
for some constant $C$. This equation can be integrated to give

\begin{equation}
k = \frac{Cw+D}{1-w^{2}}\, ,
\end{equation}

\noindent
for some other integration constant $D$. The standard form of the Dirac
monopole is recovered when we choose $C=D=p/2$. Then, Eq.~(\ref{eq:proto5})
becomes

\begin{equation}
\label{eq:proto5-2}
\Phi^{0\, \prime}+C\frac{V^{-1}}{r^{2}}
+g\xi r\Phi^{0}f
=
0\, ,  
\end{equation}

\noindent
and we are left with a non-autonomous system of 4 ordinary differential
equations for 4 variables $f,h,V,\Phi^{0}$ that generalizes Protogenov's
\cite{Protogenov:1977tq}.

The next step is to try to rewrite this system as an autonomous system by a
change of variables. For the Protogenov system this is explained in
Ref.~\cite{Meessen:2008kb}. Actually, the same change of variables works
here. Defining

\begin{equation}
gr^{2}\equiv e^{2\eta}\, ,
\hspace{.5cm}
1+gr^{2}h\equiv N\, ,
\hspace{.5cm}
gr^{2}f \equiv I\, ,
\hspace{.5cm}
gr^{2}(\Phi^{0})^{2} \equiv K^{2}\, ,
\hspace{.5cm}
C^{\prime}=g^{1/2}C\, ,
\end{equation}

\noindent
and combining the differential equations we arrive at the autonomous system

\begin{eqnarray}
\label{eq:N} 
\partial_{\eta}N
& = &
V\left[IN -\tfrac{1}{2}\xi V I^{2} +\tfrac{1}{2}g\xi V K^{2}\right]\, ,
\\
& & \nonumber \\
\label{eq:I} 
\partial_{\eta}I
& = &
(N^{2}-1)V^{-1} +I -\tfrac{1}{2}\xi V I^{2} -\tfrac{1}{2}g\xi V K^{2}\, ,
\\
& & \nonumber \\
\label{eq:V}
\partial_{\eta}V^{-1}
& = &
-\xi (N-1)V^{-1} +\xi I\, ,
\\
& & \nonumber \\
\label{eq:K}
\partial_{\eta}K
& = &
K -C^{\prime}V^{-1}-\xi V KI\, .
\end{eqnarray}

When $\xi=0$, the third equation is solved by $V=\mbox{constant}$ and, setting
that constant to $1$, the first two equations become those of the Protogenov
system and involve only two variables: $N$ and $I$. When $\xi=1$ the four
equations are coupled in a non-trivial way and we have to make additional
assumptions in order to simplify the system and find solutions.

Observe that there are no solutions with vanishing scalars, that is, with
$I=0$: setting $I=0$ in Eqs.~(\ref{eq:N}) and (\ref{eq:I}) and combining them
to eliminate $K$ we obtain a differential equation that only involves $N$ and
can be integrated to give $N=-\tanh{\eta+\alpha}$ where $\alpha$ is some
integration constant. Then, Eq.~(\ref{eq:I}) cannot be satisfied for any real
$V$ or $K$.

A further change of variables, $\mathfrak{I}= V I$ and $\mathfrak{K}= V K$,
allows us to rewrite the system in a simpler way: 

\begin{eqnarray}
\partial_{\eta} N 
& = &  
N \mathfrak{I} - \tfrac{1}{2} \mathfrak{I}^{2} + \tfrac{1}{2} g \mathfrak{K}^{2}
\, , 
\\
& & \nonumber \\
\partial_{\eta} \mathfrak{I} 
& = &  
N^{2} -1 + N \mathfrak{I} -\tfrac{3}{2} \mathfrak{I}^{2} 
-\tfrac{1}{2} g\mathfrak{K}^{2} \, , 
\\
& & \nonumber \\
\partial_{\eta} \mathfrak{K} 
& = &  
\mathfrak{K} N - C^{\prime} - 2 \mathfrak{K} \mathfrak{I} \, , 
\\
& & \nonumber \\
\partial_{\eta} \log V 
& = &  
N- \mathfrak{I}-1 \, .
\end{eqnarray}

This system admits a solution in which $N$, $\mathfrak{I}$ and $\mathfrak{K}$
are constants: the first three equations are algebraic and the fourth is
trivial to solve). This allows us to obtain the first solution of this theory.

\subsubsection{Solution 1: AdS$_{2} \times$ S$^{2}$}

With no loss of generality we can assume $\mathfrak{I}$ to be positive, and
the solution, dependent on two constants $\mathfrak{I},v$ is given by:

\begin{equation}
\begin{array}{rcl}
C^{\prime} 
& = & 
\pm \sqrt{\frac{\mathfrak{I}}{g}} \left(3 \mathfrak{I} + \sqrt{3
    \mathfrak{I}^{2} +1} \right) \left( 3 \mathfrak{I}+2 \sqrt{3
    \mathfrak{I}^{2}+1} \right)^{\frac{1}{2}} \, ,
\\
& & \\
N 
& = &  
-\mathfrak{I}-\sqrt{3 \mathfrak{I}^{2} +1} \, ,
\\
& & \\ 
\mathfrak{K} 
& = &  
\mp \sqrt{g} \left(3 \mathfrak{I}^{2} + 2 \mathfrak{I} \sqrt{3
    \mathfrak{I}^{2} +1} \right)^{\frac{1}{2}} \, ,
\\
& & \\
V 
& = &  
v g^{-\mathfrak{I}-\frac{1}{2}-\frac{1}{2} \sqrt{3 \mathfrak{I}^{2} +1}} 
r^{-2 \mathfrak{I} -1 - \sqrt{3 \mathfrak{I}^{2} +1}} \, .
\end{array}
\end{equation}

\noindent
The physical fields are then given by:
 
\begin{equation}
\begin{array}{rcl}
d s^{2} 
& = & 
{\displaystyle\frac{v^{2}}{2 \mathfrak{I}}} 
g^{-2 \mathfrak{I} +1 - \sqrt{3 \mathfrak{I}^{2}+1}} 
\left(\mathfrak{I} + \sqrt{3 \mathfrak{I}^{2}+1} \right)^{-1} 
r^{-4 \mathfrak{I} - 2 \sqrt{3 \mathfrak{I}^{2}+1}} d t^{2} \\
& & \\
&  &
- 2 \mathfrak{I} \left( \mathfrak{I}+ \sqrt{3 \mathfrak{I}^{2}+1}\right)
{\displaystyle\frac{1}{g^{2} r^{2}}} 
\left( d r^{2} + r^{2} d \Omega_{(2)}^{2} \right) \, ,
\\
& & \\
Z^{x} 
& = & 
\pm {\displaystyle\frac{x^{x}}{g r}} 
\mathfrak{I} \left( 3 \mathfrak{I}^{2}
+2 \mathfrak{I} \sqrt{3 \mathfrak{I}^{2}+1} \right)^{\frac{1}{2}}\, , 
\\
& & \\  
\Phi^{0} 
& = & 
{\displaystyle\frac{1}{v}} 
g^{\mathfrak{I}+\frac{1}{2}+\frac{1}{2} \sqrt{3 \mathfrak{I}^{2}+1}} 
\left( 3 \mathfrak{I}^{2}+ 2 \mathfrak{I} \sqrt{3
    \mathfrak{I}^{2}+1}\right)^{\frac{1}{2}} 
r^{2 \mathfrak{I}+\sqrt{3\mathfrak{I}^{2}+1}} \, ,
\\
& & \\
A^{x}{}_{\underline{m}} 
& = &
\varepsilon^{x}{}_{mn} {\displaystyle\frac{x^{n}}{g r^{2}} }
\left(-\mathfrak{I}-1-\sqrt{3 \mathfrak{I}^{2} +1} \right)\, . 
\end{array}
\end{equation}

This metric turns out to be that of AdS$_{2} \times$ S$^{2}$ (with different
radii), independently of the value of $\mathfrak{I}$, as can be seen
performing the following change of variables,
 
\begin{equation}
\begin{array}{rcl}
\rho 
& = & 
r^{-2 \mathfrak{I} - \sqrt{3 \mathfrak{I}^{2}+1}} \, , 
\\
& & \\
\tau 
& = &
v \left( 7 \mathfrak{I}^{2}+1 + 4 \mathfrak{I} \sqrt{3 \mathfrak{I}^{2}+1}\right)^{-1} 
g^{-2 \mathfrak{I}-1-\sqrt{3 \mathfrak{I}^{2}+1}}\, t  \, ,
\end{array}
 \end{equation}

\noindent
which leads to:
 
\begin{equation}
\begin{array}{rcl}
d s^{2} 
& = & 
{\displaystyle
\frac{1}{2 \mathfrak{I}} \frac{7 \mathfrak{I}^{2}+1+4\mathfrak{I}\sqrt{3
    \mathfrak{I}^{2}+1}}{\mathfrak{I}+\sqrt{3\mathfrak{I}^{2}+1}} g^{2}
\rho^{2}  d \tau^{2} 
-2\mathfrak{I} \frac{\mathfrak{I}+\sqrt{3\mathfrak{I}^{2}+1}}{7
  \mathfrak{I}^{2}+1+4\mathfrak{I}\sqrt{3 \mathfrak{I}^{2}+1}} g^{-2} \frac{d
  \rho^{2}}{\rho^{2}}
}  
\\
& & \\
&  &
- 2 \mathfrak{I} \left(\mathfrak{I}+\sqrt{3\mathfrak{I}^{2}+1}\right) g^{-2} d
\Omega_{(2)}^{2} \, ,
\\
& & \\
Z^{i} 
& = & 
\pm {\displaystyle\frac{x^{i}}{g}} 
\mathfrak{I} \left( 3 \mathfrak{I}^{2}+2 \mathfrak{I}
  \sqrt{3 \mathfrak{I}^{2}+1} \right)^{\frac{1}{2}}
\rho^{\frac{1}{2\mathfrak{I}+\sqrt{3\mathfrak{I}^{2}+1}}}\, , 
\\
& & \\
\Phi^{0} 
& = & 
{\displaystyle\frac{1}{v \rho}}
 g^{\mathfrak{I}+\frac{1}{2}+\frac{1}{2} \sqrt{3
    \mathfrak{I}^{2}+1}} \left( 3 \mathfrak{I}^{2}+ 2 \mathfrak{I} \sqrt{3
    \mathfrak{I}^{2}+1}\right)^{\frac{1}{2}} \, ,
\\
& & \\
A^{x}{}_{\underline{m}} 
& = &
\varepsilon^{x}{}_{mn} {\displaystyle\frac{x^{n}}{g}} 
\left(-\mathfrak{I}-1-\sqrt{3
    \mathfrak{I}^{2} +1} \right) \rho^{\frac{2}{2\mathfrak{I}+\sqrt{3\mathfrak{I}^{2}+1}}} \, . 
\end{array}
\end{equation}
 
 The potential (\ref{eq:potgen2cp3}) assumes in this situation a constant
 value, which can be negative for certain values of the parameter
 $\mathfrak{I}$:
 
\begin{equation}
\mathbf{V}<0
\,\,\,\,
 \Leftrightarrow 
\,\,\,\,
\mathfrak{I}^{2} \left( 3 \mathfrak{I}^{2}
  +2\mathfrak{I}\sqrt{3\mathfrak{I}^{2}+1}\right) 
< g^{2} < 
\tfrac{5}{3}\mathfrak{I}^{2} \left( 3 \mathfrak{I}^{2} 
+2\mathfrak{I}\sqrt{3\mathfrak{I}^{2}+1}\right)\, .
\end{equation}
 
By construction this solution is supersymmetric. In order to determine which
fraction of the total supersymmetry it preserves (the minimal amount is
$\frac{1}{8}$) we take advantage of the analysis performed in
Ref.~\cite{Meessen:2012sr}: the gaugini Killing Spinor Equation is solved imposing
three projection operators, each of which projects out half of the components
of the Killing spinor.  However, if some gaugini's shifts

\begin{equation}
W^{i x} = g \mathcal{G}^{i j^*} f^{* \Lambda}{}_{j^*}
\mathsf{P}_{\Lambda}{}^{x}\, ,
\end{equation}

\noindent
vanish identically for the configuration we are examining, the corresponding
projector does not need to be imposed, and the supersymmetry preserved can be
larger. From Eqs.~(\ref{eq:Kcpn}) and (\ref{eq:Vcpn}) we get, for the model we
are dealing with,

\begin{equation}
W^{i x}=0
\,\,\,\,\, 
\Leftrightarrow 
\,\,\,\,\,
Z^{i} Z^{*\, x} - \tfrac{1}{2} \delta^{i x} = 0\, ,
\end{equation}

\noindent
which can never be satisfied for the solution we are presenting, where
$Z^{x}\propto x^{x}$. This solution, therefore, is only $\frac{1}{8}$-BPS.
 

\subsection{Another ansatz}

In order to generalize the ansatz we made in Section~\ref{sec-hedgehog} we
are going to relax Eq.~(\ref{eq:HAV}): it will have the same form

\begin{equation}
 V^{x}{}_{\underline{m}} = \delta^{x}{}_{\underline{m}} V \, ,
\end{equation}

\noindent
but now we will allow $V$ to be an arbitrary (that is: not necessarily
spherically-symmetric) function of the coordinates $x^{\underline{m}}$. 

With this choice, Eq.~(\ref{eq:cp3eq3}) can be solved by 

\begin{eqnarray}
\label{eq:ansatz2A}
A^{x}{}_{\underline{m}} 
& = &
\varepsilon^{x}{}_{\underline{m} \underline{n}} h^{n} \, ,
\\
& & \nonumber \\
\label{eq:ansatz2A-2}
\partial_{\underline{m}} V 
& = &
g V \left( h^{m} + V \Phi^{m} \right) \, 
 \end{eqnarray}

\noindent
for some triplet of arbitrary functions $h^{m}$ that, in particular, can
vanish identically. We consider first this possibility.

\subsubsection{Solution 2}

Let us consider the ansatz~(\ref{eq:ansatz2A}),(\ref{eq:ansatz2A-2}) making
some further assumptions: $h^{m}=0$ and all the functions involved depend on a
single direction, say $x^{1}$, so that

\begin{equation}
\label{eq:ansatz2V}
A^{x}{}_{\underline{m}} =0 \, ,
\hspace{1cm}
\partial_{\underline{1}} V^{-1} = - g \Phi^{1} \, , 
\hspace{1cm}
\Phi^{2}=\Phi^{3}=0\, .
\end{equation}
 
\noindent
This ansatz is adequate to find domain-wall-type solutions.

Under these assumptions, Eq.~(\ref{eq:cp3eq1}) implies that the only
non-trivial component of $F^{0}{}_{mn}$ is $F^{0}{}_{23}$. However, since, by
assumption, the components $A^{0}{}_{\underline{2},\underline{3}}$ are
functions of $x^{1}$ only, they must be constants and the purely spatial
components of the field strength $F^{0}{}_{\underline{m}\underline{n}}$ must
vanish identically.

The equations in  (\ref{eq:cp3eq1}) and (\ref{eq:cp3eq2}) that remain to be
solved  are

\begin{eqnarray}
\partial_{\underline{1}} V^{-1} 
& = &
-g \Phi^{1} \, ,
\\
& & \nonumber \\
\partial_{\underline{1}} \Phi^{1} 
& = &
\tfrac{1}{2} g V \left[ (\Phi^{0})^{2}+(\Phi^{1})^{2}\right] \, ,
\\
& & \nonumber \\
\partial_{\underline{1}} \Phi^{0} 
& = & 
g \Phi^{0} \Phi^{1} V \, ,
\end{eqnarray}

\noindent
and can be rewritten in this form

\begin{eqnarray}
\partial_{V^{-1}} \Phi^{0} 
& = &
- \Phi^{0} V \, , 
\\
& & \nonumber \\
\partial_{V^{-1}} \Phi^{1} 
& = &
-\tfrac{1}{2} \frac{V}{\Phi^{1}} 
\left[ (\Phi^{0})^{2}+(\Phi^{1})^{2}\right] \, , 
\\
& & \nonumber \\
\partial_{\underline{1}} V^{-1} 
& = &
- g \Phi^{1} \, ,
\end{eqnarray}

\noindent
that can be immediately integrated, giving

\begin{equation}
\begin{array}{rcl}
\Phi^{0} 
& = & 
p^{0} V \, ,
\\ 
& & \\
\Phi^{1} 
& = & 
\pm \sqrt{\left(p^{0}\right)^{2} V^{2} + p^{1} V} \, , 
\\
& & \\
V 
& = &
-2^{\frac{5}{3}} (p^{0})^{2} (p^{1})^{2} 
\left\{
(p^{1})^{3} 
\left[
16 (p^{0})^{2}-9(p^{1})^{4} \left(-g x^{1} +v\right)^{2} 
\right]^{2} 
\right. \\
& & \\
& &
\left.
+3\sqrt{(p^{1})^{10} \left(-g x^{1}+v\right)^{2} 
\left[
-16(p^{0})^{2}+9(p^{1})^{4} \left(-g x^{1} +v\right)^{2} 
\right]^{3}
}
\right\}^{-\frac{1}{3}}\\
& & \\
& &
- 2^{\frac{1}{3}} 
\left\{
(p^{1})^{3} 
\left[16 (p^{0})^{2}-9(p^{1})^{4} \left(-g x^{1} +v\right)^{2} \right]^{2} 
\right.
\\
& & \\
& & 
\left.
+ 3 \sqrt{
(p^{1})^{10} \left(-g x^{1}+v\right)^{2} 
\left[-16 (p^{0})^{2}+9(p^{1})^{4} \left(-g x^{1} +v\right)^{2}
  \right]^{3}
}
\right\}^{\frac{1}{3}} 
\\
& & \\
& &
\left[16 (p^{0})^{2}-9(p^{1})^{4} \left(-g x^{1} +v\right)^{2} \right]^{-1}
\end{array}
\end{equation}

\noindent
where $p^{0}$, $p^{1}$ and $v$ are integration constants. The metric function
for these solutions is $e^{-2U}=(\Phi^{0})^{2}-(\Phi^{1})^{2}=-p^{1}V(x^{1})$
and the complete metric has the form

\begin{equation}
ds^{2} = -\frac{1}{p^{1}V}dt^{2} +
p^{1}V^{3}[(dx^{1})^{2}+(dx^{2})^{2}+(dx^{3})^{2}] \, .
\end{equation}

\noindent
We must set $p^{0}\neq 0$ because, otherwise, $\Phi^{0}=0$ and the metric
function would always be negative and we must require $p^{1}V<0$ so
$e^{-2U}>0$. The profile of $e^{-2U}$ changes dramatically with the
integration constants and it is not easy to find physically meaningful
solutions. One of the few simple examples that we have found corresponds to
the choice, $p^{0}=-1$, $p^{1}=1$, $v=0$, (if $g=1$) for which
$e^{-2U}(x^{1})$ is positive in an interval of the real line (see the figure
where we have represented the inverse, $e^{2U}$).
\begin{figure}[ht]
 \includegraphics[scale=0.5]{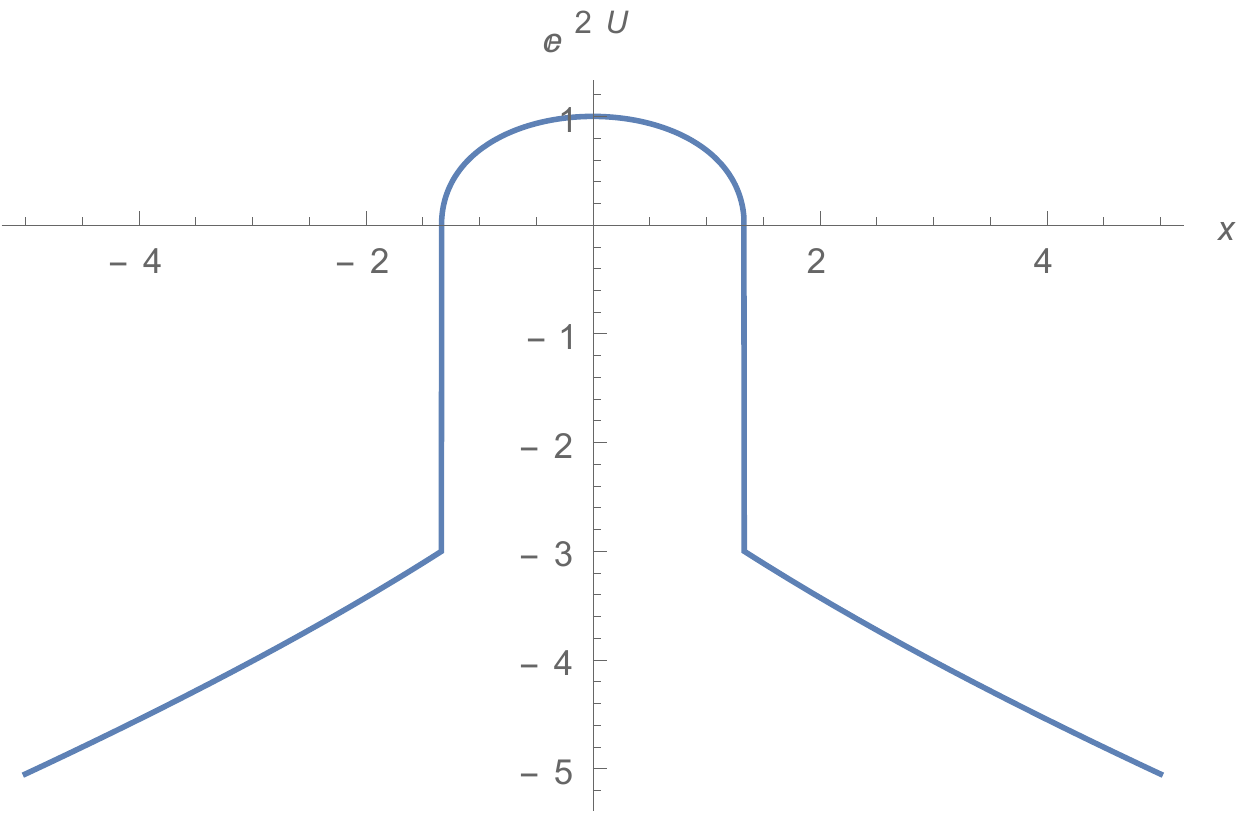}
\end{figure}

At the boundary of that region $e^{-2U}$ and $V$ blow up, and so does the
scalar potential, which in this case is given by

\begin{equation}
 \mathbf{V}=\tfrac{1}{2} g^{2} \left( 5-\frac{2 (p^{0})^{2} V}{p^{1}}\right)\, .
\end{equation}

On the other hand, the condition $W^{i x}=0$ cannot be satisfied for any $x$,
meaning that the solution is $\frac{1}{8}$-BPS.

\subsubsection{Solution 3}

If, in the context of the ansatz
Eqs.~(\ref{eq:ansatz2A}),(\ref{eq:ansatz2A-2}), we still assume that all the
functions involved depend only on $x^{1}$ but we do not assume the vanishing
of $h^{m}$, the non-trivial components of Eq.~(\ref{eq:cp3eq3}) take the form

\begin{eqnarray}
\partial_{\underline{1}} V 
& = & 
g V \left(h^{1}- V \Phi^{1} \right)\, , 
\\
& & \nonumber \\
h^{2,3}
& = &
-V \Phi^{2,3} \, ,
\end{eqnarray}
 
\noindent
those of Eq.~(\ref{eq:cp3eq2}) take the form 
 
\begin{eqnarray}
\partial_{\underline{1}} A^{0}{}_{\underline{2}} 
& = & 
- g V \Phi^{0} \Phi^{3} \, , 
\\
& & \nonumber \\
\partial_{\underline{1}} A^{0}{}_{\underline{3}} 
& = &  
g V \Phi^{0} \Phi^{2} \, , 
\\
& & \nonumber \\
\partial_{\underline{1}} \Phi^{0} 
& = & 
g V \Phi^{0} \Phi^{1} \, ,
\end{eqnarray}

\noindent
and, finally, those of Eq.~(\ref{eq:cp3eq1}) take the form
 
\begin{eqnarray}
\partial_{\underline{1}} \Phi^{2,3} 
& = & 
g h^{1} \Phi^{2,3} \, ,
\\
& & \nonumber \\
\Phi^{2} \Phi^{3} 
& = &
0\, ,
\\
& & \nonumber \\
\partial_{\underline{1}} h^{1} 
& = & 
- g V h^{1} \Phi^{1} + \tfrac{1}{2} g V^{2} 
\left[ (\Phi^{0})^{2}-(\Phi^{1})^{2}
-(\Phi^{2})^{2}+(\Phi^{3})^{2}\right]\, ,
\\
& & \nonumber \\
\partial_{\underline{1}} h^{1} 
 & = &  
- g V h^{1} \Phi^{1} + \tfrac{1}{2} g V^{2} 
\left[ (\Phi^{0})^{2}-(\Phi^{1})^{2}
+(\Phi^{2})^{2}-(\Phi^{3})^{2}\right]\, ,
\\
& & \nonumber \\
 \partial_{\underline{1}} \Phi^{1} 
& = & 
-\frac{g}{V} (h^{1})^{2} 
+ 2 g V \Phi^{2} \Phi^{3} 
+\tfrac{1}{2} g V 
\left[ (\Phi^{0})^{2}+(\Phi^{1})^{2}
-(\Phi^{2})^{2}-(\Phi^{3})^{2}\right]\, .
\end{eqnarray}

It is immediate to conclude that
 
\begin{equation}
\Phi^{2}=\Phi^{3}=0\, ,
\hspace{1cm} 
A^{0}{}_{\underline{2}\, ,\,\underline{3}}=\text{const.} \, ,
\hspace{1cm} 
A^{2}=h^{1} d x^{3} \, ,
\hspace{1cm} 
A^{3}=-h^{1} d x^{2} \, .
\end{equation}

\noindent
and the equations that remain to be solved are
 
\begin{eqnarray}
\partial_{\underline{1}} V
& = &
g V \left(h^{1}-V \Phi^{1}\right) \, ,
\\
& & \nonumber \\
\partial_{\underline{1}} h^{1} 
& = & 
-g V h^{1} \Phi^{1} 
+\tfrac{1}{2} g V^{2} \left[(\Phi^{0})^{2}-(\Phi^{1})^{2} \right]\, , 
\\
& & \nonumber \\
\partial_{\underline{1}} \Phi^{0} 
& = & 
g V \Phi^{0} \Phi^{1} \, ,
\\
& & \nonumber \\
\partial_{\underline{1}} \Phi^{1} 
& = & 
-\frac{g}{V} ( h^{1})^{2}
+\tfrac{1}{2} g V \left[(\Phi^{0})^{2}+(\Phi^{1})^{2}\right]\, .
\end{eqnarray}

This system of equations can be simplified by setting $\Phi^{1}=0$; in this
way, the resulting equations
 
\begin{eqnarray}
\Phi^{0} 
& = & 
\pm \sqrt{2} \frac{h^{1}}{V}=\text{const.}\, ,
\\
& & \nonumber \\
\partial_{\underline{1}} V  
& = &  
g V h^{1}\, ,
\\
& & \nonumber \\
\partial_{\underline{1}} h^{1}  
& = &  
g (h^{1})^{2} \, ,
\end{eqnarray}

\noindent
are easy to solve, and the solution is determined by the following
non-vanishing fields:
 
\begin{eqnarray}
\Phi^{0} 
& = & 
\pm \frac{\sqrt{2}}{b}\, ,
\\ 
& & \nonumber \\
A^{3}{}_{\underline{2}} 
& = & 
-A^{2}{}_{\underline{3}} 
=  
\frac{1}{g x^{1}} \, ,
\\
& & \nonumber \\
d s^{2}  
& = &  
\frac{2}{b^{2}} d t^{2} 
-\frac{b^{4}}{2 g^{2} (x^{1})^{2}} d x^{m} d x^{m}\, ,
\end{eqnarray}

\noindent
where $b$ is an integration constant. 

The spatial part of the metric is the metric of a 3-dimensional hyperboloid in
coordinates analogous to the Poincar\'e coordinates of AdS$_{3}$\footnote{If
  we define the hyperboloid as the hypersurface
  \begin{equation}
  (X^{1})^{2}+(X^{2})^{2} +(X^{3})^{2} -(X^{4})^{2}=-1\, ,
  \end{equation}
in the $\mathbb{R}^{4}$ endowed with the metric 
  \begin{equation}
 ds^{2}= (dX^{1})^{2}+(dX^{2})^{2} +(dX^{3})^{2} -(dX^{4})^{2}\, ,
  \end{equation}
then, if we parametrize it with coordinates $x^{1},x^{2},x^{3}$
\begin{equation}
X^{1}+X^{4}\equiv -\frac{1}{x^{1}}\, ,
\hspace{1cm}
X^{2,3}\equiv \frac{x^{2,3}}{x^{1}}\, ,  
\end{equation}
the induced metric is
\begin{equation}
ds^{2}=\frac{1}{(x^{1})^{2}}dx^{m}dx^{m}\, .  
\end{equation}
} and, therefore, the complete metric has the geometry of
$\mathbb{R}\times\mathbb{H}^{3}$ and it is supported only by a non-Abelian
field whose field strength is related to the volume form of $\mathbb{H}^{3}$
by

\begin{equation}
F^{x}{}_{yz} = -g \varepsilon_{\underline{x}\underline{y}\underline{z}}\, .  
\end{equation}

\noindent
Usually, $p$-form field strengths support $p$- of $(d-p)$-dimensional
symmetric spaces. For instance, 2-form field strengths support
AdS$_{2}\times$S$^{2}$ solutions in 4 dimensions and AdS$_{2}\times$S$^{3}$ or
AdS$_{3}\times$S$^{2}$ solutions in 5 dimensions. In this sense, this solution
is exceptional and the exceptionality is related to the rank of the form and
to the dimension of the gauge group.

The potential is again equal to a positive constant when this configuration is
considered, and the amount of supersymmetry preserved by the solution is
$\frac{1}{8}$.

\section{Solutions from dimensional reduction}
\label{sec-solutionsfromreduction}
 
An alternative procedure to construct solutions of a given theory is by
dimensional reduction or oxidation of known solutions, provided that there are
theories related to the one we are interested in by these mechanisms and that
there are known solutions of them which, if they are to be dimensionally
reduced, have enough isometries.

$\mathcal{N}=2,d=4$ supergravity theories are directly related by dimensional
reduction or oxidation to other supergravity theories with 8
supercharges.\footnote{The relation with theories with different number of
  supercharges must necessarily involve truncations and constraints on the
  solutions and we will not consider them here.}  These only exist in $d\leq
6$ and, to the best of our knowledge, theories with SU$(2)$ FI gaugings have
only been studied in $\mathcal{N}=(1,0),d=6$ supergravity coupled to one
tensor multiplet and a triplet of vector multiplets in
Ref.~\cite{Cariglia:2004kk}. This theory is unique\footnote{As different from
  $d=4,5$ supergravities with 8 supercharges, in the $d=6$ case, there is only
  one model for each possible matter content.} and describes a truncation of
the Heterotic String compactified on $T^{4}$ that includes the metric
$\tilde{g}_{\tilde{\mu} \tilde{\nu}}$, a complete\footnote{That is: not
  subject to any self- or anti-self-duality (\textit{chirality}) constraints
  because it is, actually the sum of the 2-form of the supergravity multiplet
  and the 2-form of the tensor multiplet, which have opposite chiralities}
(Kalb-Ramond) 2-form $\tilde{B}_{\tilde{\mu}\tilde{\nu}}$, a real scalar
(dilaton) $\tilde{\varphi}$ and the three vector fields
$\tilde{A}^{A}_{\tilde{\mu}}$, $A=1,2,3$. The FI terms induces a simple
potential for the dilaton, and the action takes the form
\cite{Cariglia:2004kk,Cano:2016rls}

\begin{equation}
\tilde{S}=\int d^{6}\tilde{x}\sqrt{|\tilde{g}|}
\left\{
\tilde{R}+\tfrac{1}{2}(\partial \tilde{\varphi})^{2}
+\tfrac{1}{3}e^{\sqrt{2}\tilde{\varphi}}\tilde{H}^{2}
-e^{\tilde{\varphi}/\sqrt{2}}\tilde{F}^{i}\tilde{F}^{i}
-\tfrac{3}{2} \, g_{6}^{2} \, e^{-\tilde{\varphi}/\sqrt{2}}
\right\}\, ,
\label{stringaction1}
\end{equation}

\noindent
where $g_{6}$ is the 6-dimensional coupling constant.

The dimensional reduction of $\mathcal{N}=(1,0),d=6$ supergravity theories
coupled to tensor and vector multiplets on a circle has been studied and the
models of $\mathcal{N}=1,d=5$ supergravity coupled to vector multiplets they
give rise to have been determined in Ref.~\cite{Cano:2016rls}.  We can use the
results in that paper to dimensionally reduce the 6-dimensional solutions
found in Ref.~\cite{Cariglia:2004kk} to solutions of SU$(2)$ FI-gauged
$\mathcal{N}=1,d=5$ supergravity since the relation between the 6- and
5-dimensional fields of the gauged theories is the same as in the ungauged
case, as long as the gauge groups are the same in both theories. These
relations are given Appendix~\ref{app-dimred}.  The 5-dimensional model
obtained in the dimensional reduction is completely characterized by the
symmetric tensor $C_{0 r s}= \frac{1}{3!} \eta_{r s}$, $r,s=1,\dots,5$.  The
bosonic fields in this theory are the metric $\hat{g}_{\hat{\mu} \hat{\nu}}$,
the 6 gauge fields $\hat{A}^{I}{}_{\hat{\mu}}$, $I=0,\cdots,5$, 5 of which,
$\hat{A}^{r}{}_{\hat{\mu}}$, correspond to 5 vector multiplets\footnote{The
  reduction of the KR 2-form gives just 2 vector fields.}, and 5 scalar
fields. Due to the reduction procedure, $\hat{A}^{0,1,2}{}_{\hat{\mu}}$ are
Abelian fields, while $\hat{A}^{A+2}{}_{\hat{\mu}}$ are the three SU$(2)$
gauge fields in five dimensions.  The physical scalars $\hat{\phi}^{r}$ are
encoded in the scalar functions $\hat{h}^{I}$, constrained by the fundamental
relation of Real Special Geometry

\begin{equation}
C_{IJK}\hat{h}^{I}\hat{h}^{J}\hat{h}^{K}
= 
\tfrac{1}{2}\hat{h}^{0}\eta_{rs}\hat{h}^{r}\hat{h}^{s}
=
1\, .  
\end{equation}

A convenient parametrization is $\hat{\phi}^{r}=\hat{h}^{r}$ so
$\hat{h}^{0}=2/(\phi\eta \phi)\equiv \hat{\phi}^{0}$, where $\phi\eta
  \phi\equiv\hat{\phi}^{r}\eta_{rs}\hat{\phi}^{s}$. In this parametrization,
  the last 3 scalars $\hat{\phi}^{A+2}$ transform in the adjoint
  representation of SU$(2)$ and the action of the theory can be written in the
  compact form

\begin{equation}
\begin{array}{rcl}
\hat{S} 
& = &  
{\displaystyle\int} d^{5}\hat{x}\sqrt{\hat{g}}\
\biggl\{
\hat{R}
+{\textstyle\frac{3}{2}}\hat{a}_{IJ}\hat{\mathfrak{D}}_{\hat{\mu}}\hat{\phi}^{I}
\hat{\mathfrak{D}}^{\hat{\mu}}\hat{\phi}^{J}
-{\textstyle\frac{1}{4}} \hat{a}_{IJ} \hat{F}^{I\, \hat{\mu}\hat{\nu}}
\hat{F}^{J}{}_{\hat{\mu}\hat{\nu}}
- 18 \, g_{5}^{2} \left(\hat{\phi}^{0}\right)^{-1} 
\\
& & \\
& & 
+\tfrac{1}{24\sqrt{3}}\eta_{rs}
{\displaystyle\frac{\hat{\varepsilon}^{\hat{\mu}\hat{\nu}\hat{\rho}\hat{\sigma}\hat{\alpha}}}{\sqrt{\hat{g}}}}
\hat{A}^{0}{}_{\hat{\mu}}
\hat{F}^{r}{}_{\hat{\nu}\hat{\rho}}\hat{F}^{s}{}_{\hat{\sigma}\hat{\alpha}}
\biggr\}\, ,
\end{array}
\end{equation}

\noindent
where 

\begin{equation}
\hat{\mathfrak{D}}_{\hat{\mu}}\hat{\phi}^{0,1,2}
=
\partial_{\hat{\mu}}\hat{\phi}^{0,1,2}\, ,
\hspace{1cm}
\hat{\mathfrak{D}}_{\hat{\mu}}\hat{\phi}^{A+2} 
=
\partial_{\hat{\mu}}\hat{\phi}^{A+2}
-g_{5}\epsilon^{A}{}_{BC}\hat{A}^{B}{}_{\hat{\mu}}\hat{\phi}^{C+2}\, ,    
\end{equation}

\noindent
and where the non-vanishing components of the metric $a_{IJ}$  are

\begin{equation}
a_{00}
= 
\tfrac{1}{12} (\phi\eta \phi)\, ,
\hspace{1cm}
a_{rs}   
=
\frac{-2 \eta_{rs}(\phi\eta \phi) 
+4\eta_{rr^{\prime}}\hat{\phi}^{r}\eta_{ss^{\prime}}\hat{\phi}^{r}}{3(\phi\eta
\phi)^{2}}\, .
\end{equation}

Observe that the 6- and 5- dimensional gauge coupling constants are related by 

\begin{equation}
g_{6}=\sqrt{12}g_{5}\, .  
\end{equation}

The dimensional reduction of $\mathcal{N}=1,d=5$ supergravity on a circle
gives cubic models of $\mathcal{N}=2,d=4$ supergravity. Therefore, the
$\overline{\mathbb{CP}}^{3}$ model cannot be obtained in this way. The model
that actually arises in the dimensional reduction of the above 5-dimensional
model is the ST$[2,6]$ model, which is characterized by the
prepotential\footnote{More details on this theory and, in particular, on its
  relation with the toroidal compactification of the Heterotic string can be
  found in Refs.~\cite{Bueno:2014mea,Meessen:2015enl}.}

\begin{equation}
\label{eq:prepotential}
\mathcal{F}  
= 
-\tfrac{1}{3!}
\frac{d_{ijk}\mathcal{X}^{i}\mathcal{X}^{j}\mathcal{X}^{k}}{\mathcal{X}^{0}}\, ,
\end{equation}

\noindent
where $i=1,2\cdots,6$ labels the 6 vector multiplets and where the fully
symmetric tensor $d_{ijk}$ has as only non-vanishing components

\begin{equation}
d_{1\alpha\beta}= \eta_{\alpha\beta}\, ,
\,\,\,\,\,
\mbox{where}
\,\,\,\,\,
(\eta_{\alpha\beta}) = \mathrm{diag}(+-\dotsm -)\, ,
\,\,\,\,\,
\mbox{and}
\,\,\,\,\,
\alpha,\beta=2,\cdots,6\, .
\end{equation}

\noindent
The 6 complex scalars parametrize the coset space 

\begin{equation}
\frac{\mathrm{SL}(2,\mathbb{R})}{\mathrm{SO}(2)}
\times
\frac{\mathrm{SO}(2,5)}{\mathrm{SO}(2)\times \mathrm{SO}(5)}\, , 
\end{equation}

\noindent
and the group SO$(3)$ acts in the adjoint on the coordinates $\alpha=4,5,6$
that we are going to denote with $A,B,\ldots$ indices. These are the
directions which are gauged. With our conventions, the
$\frac{\mathrm{SL}(2,\mathbb{R})}{\mathrm{SO}(2)}$ factor is parametrized by
the scalar $Z^{1}$ which is often called the axidilaton field since its real
and imaginary parts are, respectively, an axion and a dilaton field.

The action of the ST$[2,6]$ model can be constructed using the standard
formulae valid for any cubic model.\footnote{See, for instance,
  Ref.~\cite{Ortin:2015hya}.} It has a complicated form that we are not going
to use directly and, therefore, we refrain from writing it here. The
computation of the scalar potential using the general formula
Eq.~(\ref{eq:potgen2}) requires the computation of the momentum maps etc., but
we can also obtain it by dimensional reduction using the relation between 5-
and 4-dimensional fields that can be found, for instance, in
Ref.~\cite{Meessen:2015enl}. It takes the extremely simple form

\begin{equation}
\mathbf{V}(Z,Z^{*}) 
=
-\tfrac{3}{4}\, g_{4}^{2} \, \frac{1}{\Im\mathfrak{m} Z^{1}}\, ,
\end{equation}

\noindent
(that is: proportional to the exponential of the dilaton field and, therefore,
negative definite) where now the 5- and 4- coupling constants are related by

 \begin{equation}
 g_{5}= -\tfrac{1}{\sqrt{24}}\, g_{4}\, .
\end{equation}

Thus, to summarize this discussion, we can obtain supersymmetric solutions of
the above SU$(2)$ FI-gauged supergravities by dimensional reduction of the
6-dimensional supersymmetric solutions constructed in
Ref.~\cite{Cariglia:2004kk}, using the relations in the Appendix. In the rest
of this section we are going to do just that for some of those 6-dimensional
solutions.

\subsection{Solution 1}

The first solution of Ref.~ \cite{Cariglia:2004kk} that we are going to reduce
to 4 dimensions is given in Section~6.2.1 of that reference and it is,
perhaps, the simplest: it is a generalization of the solution with geometry
$\mathbb{M}_{4}\times$S$^{2}$ found by Salam in Sezgin in
Ref.~\cite{Salam:1984cj} that has $\mathbb{M}_{3}\times$S$^{3}$ metric, a
constant dilaton field whose value is proportional to the square of the radius
of the S$^{3}$ and to the square of the coupling constant, a meronic gauge
field and vanishing 2-form. The non-vanishing field are given by 

\begin{equation}
\begin{array}{rcl}
d \tilde{s}^{2} 
& = & 
d t^{2}-d z^{2} -d y^{2} - a^{2} d \Omega_{(3)}^{2} \, ,
\\
& & \\
e^{\frac{\tilde{\varphi}}{\sqrt{2}}} 
& = &
{\displaystyle
\frac{a^{2} \, g_{6}^{2}}{2}
} \, ,
\\
& & \\
\tilde{A}^{A}
& = &
{\displaystyle
-\frac{1}{2 g_{6}} \sigma^{A} 
}\, ,
\\
\end{array}
\end{equation}
 
\noindent
where the $\sigma^{A}$ are the left-invariant Maurer-Cartan 1-forms satisfying
$d \sigma^{A} = \tfrac{1}{2} \varepsilon^{A}_{BC} \sigma^{B} \wedge
\sigma^{C}$, $d \Omega_{(3)}^{2} = \frac{1}{4} \sigma^{A} \sigma^{A}$ and $a$
is a constant parameter.

Reducing along the $z$ coordinate using Eqs.~(\ref{eq:6to5}), we get a
solution of the 5-dimensional theory with the following non-vanishing fields:
 
\begin{equation}
\begin{array}{rcl}
d \hat{s}^{2}  
& = &  
d t^{2}-d y^{2}-a^{2} d \Omega_{(3)}^{2} \, ,
\\
& & \\
\hat{h}^{0} 
& = & 
6 a^{2} g_{5}^{2} \, , 
\\
& & \\
\hat{h}^{1} 
& = & 
1+{\displaystyle\frac{1}{12 a^{2} g_{5}^{2}}} \, , 
\\
& & \\
\hat{h}^{2} 
& = & 
1-{\displaystyle\frac{1}{12 a^{2} g_{5}^{2}}} \, , 
\\
& & \\
\hat{A}^{A+2} 
& = & 
-{\displaystyle\frac{1}{2 g_{5}}} \sigma^{A} \, .
\end{array}
\end{equation}

This solution belongs to the same class as its 6-dimensional parent: it has
constant scalars and a meronic gauge field that support an
$\mathbb{M}_{2}\times$S$^{3}$ geometry.

Reducing further along the $y$ coordinate using Eqs.~(\ref{eq:5to4}), we
obtain a 4-dimensional solution of the same kind with non-vanishing fields
 
\begin{equation}
\begin{array}{rcl}
d s^{2} 
& = & 
d t^{2}- a^{2} d \Omega_{(3)}^{2} \, ,
\\
& & \\
Z^{1} 
& = &  
\frac{i}{4}\, a^{2} g_{4}^{2} \, ,
\\
& & \\
Z^{2} 
& = &  
i \left(1+{\displaystyle\frac{2}{a^{2} g_{4}^{2}}} \right) \, ,
\\
& & \\
Z^{3} 
& = &  
i \left(1-{\displaystyle\frac{2}{a^{2} g_{4}^{2}}} \right) \, ,
\\
& & \\
A^{A+3} 
& = & 
-{\displaystyle\frac{1}{2 \, g_{4}}} \sigma^{A}\, .
\end{array}
 \end{equation}

\subsection{Solution 2}

The second solution we are going to consider is the \textit{dyomeronic black
  string} constructed in Section~6.2.2 of Ref.~\cite{Cariglia:2004kk}, which
corresponds to a black string lying along the $z$ direction with electric and
magnetic 3-form and a meronic gauge field in the 4-dimensional
transverse space.  Its non-vanishing fields are given by

\begin{equation}
\begin{array}{rcl}
d \tilde{s}^{2}  
& = &   
{\displaystyle\frac{r}{\sqrt{Q_{1}+\frac{Q_{2}}{r^{2}}}}} 
\left(d t^{2}-d z^{2}\right) -
{\displaystyle\frac{\sqrt{Q_{1}+\frac{Q_{2}}{r^{2}}}}{r}} 
\left(d r^{2} + a^{2} r^{2} \, d \Omega_{(3)}^{2} \right) \, , 
\\
& & \\
e^{\sqrt{2} \tilde{\varphi}}  
& = &  
{\displaystyle\frac{a^{4} g_{6}^{4}}{4 \left(1-a^{2}\right)^{2}}} 
r^{2}\left(Q_{1}+{\displaystyle\frac{Q_{2}}{r^{2}}} \right) \, ,
\\
& & \\
\tilde{A}^{i}  
& = &  
-{\displaystyle\frac{1-a^{2}}{2 g_{6}}} \sigma^{i} \, , 
\\
& & \\
\tilde{H}  
& = &  
{\displaystyle\frac{1-a^{2}}{g_{6}^{2}}} 
\left[
{\displaystyle\frac{a}{4}} r\, \sigma^{1} \wedge \sigma^{2}\wedge \sigma^{3} 
+{\displaystyle\frac{2 Q_{2}}{a^{2}}}
{\displaystyle\frac{1}{r^{3} \left(Q_{1}+\frac{Q_{2}}{r^{2}}\right)^{2}}} 
\, d t \wedge d r \wedge d z 
\right] \, .
\end{array}
\end{equation}

\noindent
where the parameter $a$ satisfies $a^{2}<1$. This solution is not
asymptotically AdS (or some other known vacuum solution) but has a horizon at
$r=0$ and in the near-horizon limit $r\rightarrow 0$ the metric is of the form
AdS$_{3}\times$S$^{3}$ where the two factors have different radii. Since this
limit is equivalent to setting $Q_{1}=0$, the AdS$_{3}\times$S$^{3}$
near-horizon limit is a supersymmetric solution as well.

If we reduce along the $z$ direction, the following 5-dimensional solution is obtained

\begin{equation}
\begin{array}{rcl}
d \hat{s}^{2}  
& = &  
{\displaystyle
r^{\frac{4}{3}} \left(Q_{1}+\frac{Q_{2}}{r^{2}}
  \right)^{-\frac{2}{3}} d t^{2} - r^{-\frac{2}{3}}
  \left(Q_{1}+\frac{Q_{2}}{r^{2}} \right)^{\frac{1}{3}} \left( d r^{2} + a^{2}
    r^{2} d \Omega_{(3)}^{2} \right) \, , 
}
\\
& & \\
\hat{h}^{0}  
& = &  
{\displaystyle
\frac{6 \,a^{2} g_{5}^{2}}{1-a^{2}} \, r^{\frac{4}{3}}
 \left(Q_{1}+\frac{Q_{2}}{r^{2}} \right)^{\frac{1}{3}} \, , 
}
\\
& & \\
\hat{h}^{1}  
& = &  
{\displaystyle
r^{-\frac{2}{3}} \left(Q_{1}+\frac{Q_{2}}{r^{2}} \right)^{\frac{1}{3}} 
\left[
 1+ \frac{1-a^{2}}{12 \, a^{2} g_{5}^{2} \left(Q_{1}+\frac{Q_{2}}{r^{2}}
\right)}
\right]\, , 
}
\\
& & \\
\hat{h}^{2}  
& = &  
{\displaystyle
r^{-\frac{2}{3}} \left(Q_{1}+\frac{Q_{2}}{r^{2}} \right)^{\frac{1}{3}} 
\left[
1- \frac{1-a^{2}}{12 \, a^{2} g_{5}^{2} \left(Q_{1}+\frac{Q_{2}}{r^{2}}
\right)}
\right] \, , 
}
\\
& & \\
\hat{F}^{0}  
& = &  
{\displaystyle
12^{2} \sqrt{3}\, g_{5}^{2} \frac{a^{2}}{1-a^{2}} \, 
r^{\frac{5}{2}} \left(Q_{1}+\frac{Q_{2}}{r^{2}} \right)^{-\frac{1}{4}} d t
\wedge d r \, ,
}
\\
& & \\
\hat{F}^{1}  
& = & 
{\displaystyle
-\hat{F}^{2} =\frac{1-a^{2}}{2 \sqrt{3}\, a^{2}
    g_{5}^{2}} \, \frac{Q_{2}}{r^{3}} \left(Q_{1}+\frac{Q_{2}}{r^{2}}
  \right)^{-2} d t \wedge d r \, , 
}
\\
& & \\
\hat{A}^{A+2}  
& = &  
{\displaystyle
- \frac{1-a^{2}}{2 \, g_{5}} \, \sigma^{A} \, . 
}
\end{array}
\end{equation}

This solution is singular at $r=0$ and it is not asymptotically AdS (or some
other known vacuum solution). If we reduce it again along the coordinate
$\phi$, defined by $d \Omega_{(3)}^{2} = \frac{1}{4} \left[ \left(d \phi +
    \cos \theta\, d \psi \right)^{2} + d \theta^{2}+\sin^{2} \theta \, d
  \psi^{2} \right]$, we get a 4-dimensional solution which we will refrain
from writing explicitly because it has the same problems as the 5-dimensional
one.

Of course, we could have used this coordinate $\phi$ in the reduction from 6
to 5 dimensions. Doing that we get a 5-dimensional solution with the properties
similar to those of the 6-dimensional one:

\begin{equation}
\begin{array}{rcl}
d \hat{s}^{2}  
& = &  
{\displaystyle
\left( \frac{a}{2} \right)^{\frac{2}{3}} 
r^{\frac{4}{3}} \left(Q_{1}+\frac{Q_{2}}{r^{2}} \right)^{-\frac{1}{3}} 
\left( d t^{2} - d z^{2} \right)
-\left( \frac{a}{2} \right)^{\frac{2}{3}} 
 r^{-\frac{2}{3}} \left( Q_{1}+\frac{Q_{2}}{r^{2}}
  \right)^{\frac{2}{3}} d r^{2}
}
\\
& & \\
&  & 
{\displaystyle
-\left(\frac{a}{2}\right)^{\frac{8}{3}} r^{\frac{4}{3}} \left(
  Q_{1}+\frac{Q_{2}}{r^{2}} \right)^{\frac{2}{3}} d \Omega_{(2)}^{2} \, , 
}
\\
& & \\
\hat{h}^{0}  
& = &  
{\displaystyle
\frac{3 \cdot 2^{\frac{1}{3}} a^{\frac{8}{3}} g_{5}^{2}}{1-a^{2}} \,
r^{\frac{4}{3}} \left(Q_{1}+\frac{Q_{2}}{r^{2}}\right)^{\frac{2}{3}} \, , 
}
\\
& & \\
\hat{h}^{1}  
& = &  
{\displaystyle
\left(\frac{2}{a}\right)^{\frac{4}{3}} \, r^{-\frac{2}{3}}
\left(Q_{1}+\frac{Q_{2}}{r^{2}}\right)^{-\frac{1}{3}}
\left[1+\frac{\left(a^{2}-1\right) \left(a^{2}-2\right)}{4 \cdot 12 \,
    g_{5}^{2}}\right]\, , 
}
\\
& & \\
\hat{h}^{2}  
& = &  
{\displaystyle
\left(\frac{2}{a}\right)^{\frac{4}{3}} \, r^{-\frac{2}{3}}
\left(Q_{1}+\frac{Q_{2}}{r^{2}}\right)^{-\frac{1}{3}}
\left[1-\frac{\left(a^{2}-1\right) \left(a^{2}-2\right)}{4 \cdot 12
    \,g_{5}^{2}}\right]\, , 
}
\\
& & \\
\hat{h}^{A+2}  
& = &  
{\displaystyle
-\frac{1-a^{2}}{2^{\frac{2}{3}} \cdot 3 \,a^{\frac{4}{3}} g_{5}} \,
r^{-\frac{2}{3}}  \left( Q_{1}+\frac{Q_{2}}{r^{2}} \right)^{-\frac{1}{3}} 
\frac{x^{A}}{r}\, ,
}
\\
& & \\
\hat{F}^{0} 
& = & 
{\displaystyle
\frac{3^{\frac{5}{2}} a^6 g_{5}^{2} }{1-a^{2}}\, Q_{2}\, r^{\frac{3}{2}}
\left( Q_{1}+\frac{Q_{2}}{r^{2}} \right)^{-\frac{1}{4}} \cos \theta \, d
\theta \wedge d \psi \, , 
}
\\
& & \\
\hat{F}^{1}  
& = &  
-\hat{F}^{2}   
=   
{\displaystyle
\left[\frac{\left(1-a^{2}\right) a}{16 \sqrt{3} \, g_{5}^{2}} \, r -2 \sqrt{3}
\right] \sin \theta \,d \theta \wedge d \psi\, , 
}
\\
& & \\
\hat{A}^{3}  
& = &  
{\displaystyle
\frac{1-a^{2}}{2 \,g_{5}} \left( -\sin \psi \,d \theta +\cos \theta \sin\theta \cos \psi \,d \psi \right) \, , 
}
\\
& & \\
\hat{A}^{4}  
& = &  
{\displaystyle
\frac{1-a^{2}}{2\, g_{5}} \left( \cos \psi\, d \theta +\cos \theta \sin \theta
  \sin \psi\, d \psi \right) \, , 
}
\\
& & \\
\hat{A}^5  
& = &  
{\displaystyle
-\frac{1-a^{2}}{2 \,g_{5}} \cos \theta \left(1+\cos \theta \right) d \psi \, ,
}
\end{array}
\end{equation}

\noindent
where we have introduced 3 Cartesian coordinates $x^{A}$ related to the
spherical coordinates $r,\theta,\psi$ in the standard way.

This solution is regular in the $r\rightarrow 0$ limit, where the metric
becomes that of the product AdS$_{3} \times$ S$^{2}$ with different radii:

\begin{equation}
d\hat{s}^{2} 
\rightarrow 
\left(\frac{a}{2}\right)^{2/3} \frac{Q_{2}^{2/3}}{\rho^{2}} 
\left(d t^{2}- dz^{2}-d\rho^{2}\right) -\left(\frac{a}{2}\right)^{8/3} 
Q_{2}^{\frac{2}{3}} \,d \Omega_{2}^{2} \, .
\end{equation}

\noindent
where $\rho\equiv Q_{2}^{1/2}/r$ but, again, it is not asymptotically
AdS. 

The $r\rightarrow 0$ limit of the complete solution coincides with the
solution that one gets by setting $Q_{1}=0$. Thus, there is a globally regular
AdS$_{3} \times$ S$^{2}$ solution in this theory. It could have been obtained
directly by dimensional reduction from the 6-dimensional AdS$_{3} \times$
S$^{3}$ solution.

Further reduction along the $z$ coordinate would lead to the same
4-dimensional solution mentioned above. There are, however, other possibilities
inspired in the results of Ref.~\cite{LozanoTellechea:2002pn}, in which the
relation between AdS$_{n} \times$ S$^{m}$ vacua of the 4-, 5- and
6-dimensional theories with 8 supercharges was studied. The main observation
is that, just as S$^{3}$ can be seen as a U$(1)$ fibration over S$^{2}$ and
one gets that S$^{2}$ by dimensional reduction along that fiber\footnote{This
  is what we have done here to go from the AdS$_{3} \times$ S$^{3}$ to the
  AdS$_{3} \times$ S$^{2}$ solution.}, AdS$_{3}$ can be seen as a U$(1)$
fibration over AdS$_{2}$ and, by dimensional reduction along that fiber one
gets AdS$_{2}$. Thus, if instead of using the coordinate $z$ along which the
6-dimensional string lies, one uses the U$(1)$ fiber of the AdS$_{3}$ in the
AdS$_{3} \times$ S$^{3}$ solution, we would have obtained an AdS$_{2} \times$
S$^{3}$ solution in 5 dimensions and then an AdS$_{2} \times$ S$^{2}$ solution
in 4 dimensions.

A more general dimensional reduction is possible: one can rotate the two
U$(1)$ fibers of the 6-dimensional solution and dimensionally reduce along one
of the rotated fibers. As in the ungauged case studied in
Ref.~\cite{LozanoTellechea:2002pn} one would get a solution that describes
geometry of the near-horizon limit of the BMPV black hole in which the
remaining U$(1)$ is non-trivially fibered over AdS$_{2} \times$S$^{2}$. This
space is obtained in 4 dimensions after dimensional reduction along the
remaining fiber.

The main difference with the ungauged case, apart from the presence of
non-trivial SU$(2)$ gauge field, is the difference between the radii of the
two factors of these metrics.

Carrying out these alternative dimensional reductions following
Ref.~\cite{LozanoTellechea:2002pn} is straightforward, albeit quite involved
due to the necessity to rewrite the 6-dimensional solution in different
coordinates. We leave it for a future publication.

\section{Conclusions}
\label{sec-conclusions}

Exploring the space of the supersymmetric solutions of a supergravity theory
is one of the most elementary steps one can take to get a more complete
understanding of its structure, providing information about the possible vacua
and some of the solitonic objects that can exist on it. In this paper we have
taken this step for two particular examples (the $\overline{\mathbb{CP}}^{3}$
and ST$[2,6]$ models) of a wide class of theories with a class of gaugings
that has been overlooked so far: SU$(2)$-FI-gauged $\mathcal{N}=2,d=4$
supergravity.

Although, as we have shown, no maximally supersymmetric solutions exist in
these theories, there are non-maximally-supersymmetric solutions that can be
seen as a deformation of the maximally supersymmetric vacua of the ungauged
theory, such as the AdS$_{2}\times$S$^{2}$ solutions with different radii.
Actually, in the ST$[2,6]$ model, the AdS$_{2}\times$S$^{2}$ solution must
have a higher-dimensional origin analogous to that of the ungauged case
\cite{LozanoTellechea:2002pn} and we have indicated the existence of a family
of vacua of $\mathcal{N}=1,d=5$ supergravity similar to the near-horizon
geometry of the 5-dimensional BMPV black-hole originating in the
AdS$_{3}\times$S$^{3}$ solution of $\mathcal{N}=(1,0),d=6$ supergravity with
different radii constructed by Cariglia and Mac Conamhna in
Ref.~\cite{Cariglia:2004kk}.

It is likely the existence of deformed versions of the rest of the maximally
supersymmetric vacua of $\mathcal{N}=1,d=5$ supergravity (H$pp$-waves and
G\"odel spacetimes \cite{Meessen:2001vx,Gauntlett:2002nw}). It may be possible
to obtain them from the above-mentioned solutions by different limiting
procedures \cite{kn:SO}. On the other hand, it would be interesting to find
complete black-hole and black-string solutions whose near-horizon geometries
were precisely the AdS$_{m}\times$S$^{n}$ solutions we have discussed, but it
is not guaranteed that they are always going to exist and their asymptotic
behaviour is uncertain.

Apart from these solutions we have found solutions whose geometry is of the
form $\mathbb{M}_{m}\times$ S$^{n}$ in 4 and 5 dimensions which descend from a
6-dimensional solution of the same kind and a solution of the
$\overline{\mathbb{CP}}^{3}$ model with $\mathbb{R}\times\mathbb{H}^{3}$
geometry which deserves further study. Work in this direction is in progress
\cite{kn:SO}.

\section*{Acknowledgments}

C.S.~would like to thank Samuele Chimento for many very useful conversations.
This work has been supported in part by the Spanish Ministry of Science and
Education grant FPA2012-35043-C02-01, the Centro de Excelencia Severo Ochoa
Program grant SEV-2012-0249, and the Spanish Consolider-Ingenio 2010 program
CPAN CSD2007-00042.  TO wishes to thank M.M.~Fern\'andez for her permanent
support.

\appendix
\section{Rules for dimensional reduction}
\label{app-dimred}

\subsection{6 $\rightarrow$ 5}

Following Ref.~\cite{Cano:2016rls}, for the supergravity theories considered
in Section~\ref{sec-solutionsfromreduction}, if we perform the dimensional
reduction along the coordinate $z$, the 5-dimensional fields of can be
expressed in terms of the 6-dimensional fields ones as follows:

\begin{equation}
\label{eq:6to5}
\begin{array}{rcl}
 \hat{g}_{\hat{\mu} \hat{\nu}}  
 & = &  
\tilde{g}_{\hat{\mu} \hat{\nu}} \left| \tilde{g}_{\underline{z} \underline{z}}
\right|^{\frac{1}{3}} + \tilde{g}_{\hat{\mu} \underline{z}} \,
\tilde{g}_{\hat{\nu} \underline{z}} \left| \tilde{g}_{\underline{z}
    \underline{z}} \right|^{-\frac{2}{3}} \, ,
\\
& & \\
\hat{h}^{0}  
& = &  
e^{\frac{\tilde{\varphi}}{\sqrt{2}}} \left| \tilde{g}_{\underline{z}
    \underline{z}} \right|^{\frac{1}{3}} \, , 
\\
& & \\
\hat{h}^{1}  
& = &  
\left| \tilde{g}_{\underline{z} \underline{z}} \right|^{-\frac{2}{3}}
\left(1+\tilde{A}^{i}{}_{\underline{z}} \tilde{A}^{i}{}_{\underline{z}} \right)
+\frac{1}{2} e^{-\frac{\tilde{\varphi}}{\sqrt{2}}} \left|
  \tilde{g}_{\underline{z} \underline{z}} \right|^{\frac{1}{3}} \, ,
\\
& & \\
\hat{h}^{2}  
& = &  
\left| \tilde{g}_{\underline{z} \underline{z}} \right|^{-\frac{2}{3}}
\left(1-\tilde{A}^{i}{}_{\underline{z}} \tilde{A}^{i}{}_{\underline{z}} \right)
-\frac{1}{2} e^{-\frac{\tilde{\varphi}}{\sqrt{2}}} \left|
  \tilde{g}_{\underline{z} \underline{z}} \right|^{\frac{1}{3}} \, ,
\\
& & \\
\hat{h}^{i+2}  
& = &  
- 2 \left| \tilde{g}_{\underline{z} \underline{z}} \right|^{-\frac{2}{3}}
\tilde{A}^{i}{}_{\underline{z}} \, , 
\\
& & \\
\hat{F}^{0}{}_{\hat{a} \hat{b}}  
& = & 
-4 \sqrt{3} \left| \tilde{g}_{\underline{z} \underline{z}}
\right|^{\frac{2}{3}} e^{\sqrt{2} \tilde{\varphi}} \, \epsilon_{\hat{a}
  \hat{b} \hat{c} \hat{d} \hat{e}} \, \tilde{H}^{\hat{c} \hat{d} \hat{e}} \, , 
\\
& & \\
\hat{F}^{1}{}_{\hat{\mu} \hat{\nu}}   
& = &  
\sqrt{3}\, \tilde{H}_{\hat{\mu} \hat{\nu} \underline{z}} 
+4 \sqrt{3}\, \tilde{A}^{i}{}_{\underline{z}}\,
\tilde{F}^{i}{}_{\hat{\mu}\hat{\nu}} 
+ 2 \sqrt{3}\, \partial_{[\hat{\mu}} 
\left[ 
{\displaystyle
\frac{\tilde{g}_{\hat{\nu}]\underline{z}}}{\tilde{g}_{\underline{z}\underline{z}}}
} 
\left(\tilde{A}^{i}{}_{\underline{z}} \tilde{A}^{i}{}_{\underline{z}}+1\right)
\right] \, , 
\\
& & \\
\hat{F}^{2}{}_{\hat{\mu} \hat{\nu}}   
& = &  
-\sqrt{3}\, \tilde{H}_{\hat{\mu} \hat{\nu} \underline{z}} 
- 4 \sqrt{3}\, \tilde{A}^{i}{}_{\underline{z}}\,
\tilde{F}^{i}{}_{\hat{\mu}\hat{\nu}} -2 \sqrt{3}\, \partial_{[\hat{\mu}} 
\left[ 
{\displaystyle
\frac{\tilde{g}_{\hat{\nu}]\underline{z}}}{\tilde{g}_{\underline{z}\underline{z}}} }
\left(\tilde{A}^{i}{}_{\underline{z}} \tilde{A}^{i}{}_{\underline{z}}-1\right)
\right] \, , 
\\
& & \\
\hat{A}^{i+2}{}_{\hat{\mu}}  
& = & 
\sqrt{12} \, \tilde{A}^{i}{}_{\hat{\mu}}+2 \sqrt{3} \, 
{\displaystyle
\frac{\tilde{g}_{\hat{\mu}\underline{z}}}{\tilde{g}_{\underline{z}\underline{z}}}
}\, 
\tilde{A}^{i}{}_{\underline{z}} \, .
\end{array}
\end{equation}

\subsection{5 $\rightarrow$ 4}

Following Ref.~\cite{Meessen:2015enl}, for the supergravity theories
considered in Section~\ref{sec-solutionsfromreduction}, if we perform the
dimensional reduction along the coordinate $y$, the 4-dimensional fields can
be expressed in terms of the 5-dimensional ones as follows:

\begin{equation}
\label{eq:5to4}
\begin{array}{rcl}
g_{\mu \nu}  
& = &  
\left|\hat{g}_{\underline{y} \underline{y}}\right|^{\frac{1}{2}} 
\left[
\hat{g}_{\mu \nu}-\frac{\hat{g}_{\mu \underline{y}} \, 
\hat{g}_{\nu \underline{y}}}{\hat{g}_{\underline{y} \underline{y}}} 
\right] \, , 
\\
& & \\
Z^{i} 
& = & 
\frac{1}{\sqrt{3}} \hat{A}^{i-1}{}_{\underline{y}} 
+i\left|\hat{g}{}_{\underline{y} \underline{y}}\right|^{\frac{1}{2}}
\hat{h}^{i-1} \, ,
\\
& & \\
A^{0}{}_{\mu}  
& = &  
\frac{1}{2 \sqrt{2}} 
{\displaystyle
\frac{\hat{g}_{\mu \underline{y}}}{\hat{g}_{\underline{y}\underline{y}}}
} \, , 
\\
A^{i}{}_{\mu}  
& = &  
-\frac{1}{2 \sqrt{6}} 
\left[
\hat{A}^{i-1}{}_{\mu} - \hat{A}^{i-1}{}_{\underline{y}} \, 
{\displaystyle
\frac{\hat{g}_{\mu \underline{y}}}{\hat{g}_{\underline{y} \underline{y}}}
}
\right] \, .
\end{array}
\end{equation}


\end{document}